\documentclass{elsart}
\newcommand{\bra}[1]{\left\langle #1\right|}
\newcommand{\ket}[1]{\left|#1\right\rangle}
\newcommand{\psibeta}[0]{\psi^{\lambda}}
\newcommand{\ketbra}[0]{\ket{\psibeta}\bra{\psibeta}}
\newcommand{\lambdabeta}[0]{\lambda}

\usepackage{amsmath,amssymb,epsfig}
\usepackage{latexsym,hyperref}
\usepackage{graphicx}
\usepackage{algorithmic}

\newcommand{\sn}{\text{sn}}

\def\sign#1{\hbox{\rm \,sign}(#1)}

\newcommand{\eq}[1]{(\ref{#1})}

\makeatother

\newcommand{\eqnref}[1]{(\ref{#1})}

\begin{document}

\runauthor{N.\ Cundy et al.}
\begin{frontmatter}



\title{Numerical Methods for the QCD Overlap Operator: 
       III. Nested Iterations}


%
%
\author[Wupphys]{N. Cundy},
\author[Ddorf]{J.\ van den Eshof},
\author[Wupmat]{A.\ Frommer},
\author[Wupphys]{S. Krieg},
\author[ZAM]{Th. Lippert} and
\author[Wupmat]{K. Sch\"afer}

\address[Wupphys]{Department of Physics, University of Wuppertal,
Germany}

\address[Ddorf]{Department of Mathematics, University of D\"usseldorf,
Germany}

\address[Wupmat]{Department of Mathematics, University of Wuppertal,
Germany}

\address[ZAM]{Central Institute for Applied Mathematics, Research Center J\"ulich,
Germany}

\begin{abstract}
  The numerical and computational aspects of chiral fermions in
  lattice quantum chromodynamics are extremely demanding. In the
  overlap framework, the computation of the fermion propagator
  leads to a nested iteration where the matrix vector
  multiplications in each step of an outer iteration have to be
  accomplished by an inner iteration; the latter approximates the
  product of the sign function of the hermitian Wilson fermion
  matrix with a vector.
  
  \noindent In this paper we investigate aspects of this nested paradigm. We
  examine several Krylov subspace methods to be used as an outer
  iteration for both propagator computations and the Hybrid Monte-Carlo
  scheme. We establish criteria on the accuracy of the inner
  iteration which allow to preserve an a priori given precision
  for the overall computation. It will turn out that the accuracy
  of the sign function can be relaxed as the outer iteration
  proceeds.  Furthermore, we consider preconditioning strategies,
  where the preconditioner is built upon an inaccurate
  approximation to the sign function.  Relaxation combined with
  preconditioning allows for considerable savings in computational
  efforts up to a factor of 4 as our numerical experiments
  illustrate. We also discuss the possibility of projecting the
  squared overlap operator into one chiral sector.
\end{abstract}

\end{frontmatter}


\section{Introduction}

For two decades numerical simulations of very light quarks within
lattice quantum chromodynamics have remained intractable as the
chiral symmetry of the underlying QCD Lagrangian, which holds in
the case of zero mass quarks, could not be embedded into flavour
conserving fermion lattice discretisation schemes. The standard
workaround took recourse to simulations with fairly heavy quarks
instead and extrapolated the results over a wide range of quark
masses to the very light quark mass regime.  Unfortunately,
simulating far beyond the realm of chiral perturbation theory such
extrapolations carry large systematic errors which have to be
avoided in order to achieve a sufficient precision of
phenomenological observables \cite{Namekawa:2004bi}.\phantom{\cite{Wuppertal:1999}}

It was realised by Hasenfratz some years ago \cite{Hasenfratz} that
considerable progress can be achieved in this bottleneck problem
through switching to a discretisation scheme that obeys a lattice
variant of chiral symmetry, as expressed by the Ginsparg-Wilson
relation for the quark propagator \cite{Ginsparg:1982bj} which in
turn implies a novel version of chiral symmetry on the lattice
\cite{Luscher:1998du}.  Theoretically, such a scheme induces a
dramatic reduction in fluctuations in the vicinity of quark mass
zero. Shortly before the rediscovery of the Ginsparg-Wilson relation,
Neuberger had constructed the overlap operator
\cite{Neuberger:1998fp,Narayanan:2000qx}, a very
promising candidate for a chiral Dirac operator~\cite{Neuberger:1998bg,Neuberger:1998wv}. It implies the solution of linear systems
involving the inverse
matrix square root or the matrix sign function (of the hermitian
Wilson-Dirac operator $Q$). This can be turned into
an intriguing practical method to simulate light quarks through
iterative methods following an inner-outer paradigm: One performs an
outer Krylov subspace
method where each iteration requires the computation of a matrix-vector
product involving $\sign{Q}$. Each such product is computed through
another, inner, iteration using matrix-vector multiplications with $Q$.

The problem of approximating the action of $\sign{Q}$ on a vector has
been dealt with in a number of papers, using polynomial approximations
\cite{HeJaLe99,Hernandez:1999gu,HJL00,Bu98,Hernandez:2000iw}, Lanczos
based methods \cite{Bor99c,Bor99b,Bor99a,Vor00} and multi-shift CG
combined with a partial fraction expansion \cite{Neuberger:1998my,Neu00,EHKN00,EHN98}. 
In an earlier paper \cite{EFL02} we
have introduced the Zolotarev partial fraction approximation (ZPFE) as the
optimal approximation to the matrix sign function. ZPFE has led to an
improvement of about a factor of 3 compared to the Chebyshev polynomial
approach \cite{HJL00}. This technique to compute the sign function
is meanwhile established as the method of choice,
\cite{OVERLAP1,OVERLAP2,LIUOVERLAP}. Moreover, it is the natural starting
point for both the simulation of dynamical overlap fermions \cite{FODOR,paper4}
and so-called optimised domain wall fermions \cite{LIU1,LIU2,LIULATTICE}.

So far simulations with overlap fermions have been restricted to the
quenched model, where fermion loops are neglected, because of the sheer costs
of the evaluation of the sign function on matrices with extremely high dimensions
\cite{HJL00,Jansen2,Giusti1,GHL02}. The challenge today is to step
away from the quenched model and include dynamical fermions. At this point
we have the unique opportunity to devise optimised simulation algorithms
for overlap fermions, investigating novel numerical and stochastical
techniques.

Indeed, efficient methods to compute the sign function are only
half of the story. It is equally important to design the entire
nested iteration in an optimal manner. This means that we should
care on how accurately we actually need the sign function to be
computed in each step of the outer iteration process in order to
achieve a given accuracy. As we will show, to achieve a given
accuracy for the solution of the entire system, one can relax the
accuracy of the computation of the sign function as the iteration
proceeds.  In this manner, the computational effort is reduced
substantially.  In addition to this approach, we will use the
concept of recursive preconditioning of a Krylov subspace method
to obtain further accelerations. 

In the present paper---which is part of a continuing series---we
show that the use of relaxation strategies and recursive
preconditioning in the linear system solver will substantially
improve over existing methods, gaining a factor of 3 to 4 in
computational speed in dynamical simulations on realistic
lattices. Together with the improvement of ZPFE over Chebyshev
polynomials, we
therefore now have an improvement factor of
about 10 over early overlap propagator computations
\cite{HJL00}.
These results are practical without any restrictions, i.e.,\ they rely
on available computed quantities only. We do assume that there are
computable error bounds for the approximation quality of the sign
function. As was shown in our earlier paper \cite{EFL02}, this is
the case for the Zolotarev approach using multi-shift CG (Theorem 7
in \cite{EFL02}) as well as for a Lanczos based approach for $Q^2$
(Theorem 6 in \cite{EFL02}).
All our results are obtained projecting out a number of low
lying eigenvectors of the hermitian Wilson fermion operator. We
briefly discuss the optimisation of the number of projected
eigenvectors, taking into account the additional
effort to generate these low lying modes by means of the
Arnoldi algorithm.

The paper is organised as follows: in Section~\ref{formsec} we
briefly review results from \cite{vdEetal:03a} which relate
different formulations of Neuberger's operator to optimal Krylov
subspace methods for the solution of the corresponding linear
systems.  In Section~\ref{inexactsec} we apply the results from
\cite{SEs02} to these methods and we obtain strategies on how to
choose the accuracy for the inner iteration (evaluating the matrix
vector multiplication $\sign{Q}y$) at each step of the outer
iteration.

Section~\ref{precsec} presents further improvements based on the
`recursive' preconditioning technique, i.e.,\ we use an inaccurate solver for
the system as a preconditioner for each step of the outer iteration. As we
will point out, recursive preconditioning might be considered a
generalisation and improvement of approaches suggested by Giusti et al.
\cite{GHL02} and Bori\c{c}i \cite{Bor99}.  For the purpose of illustration,
Sections~\ref{inexactsec} and \ref{precsec} will contain results from
numerical calculations for a realistic, but small ($4^4$), example configuration.
Results on more numerical experiments are given in
Section~\ref{numericalsec} where we achieve improvement factors in a range from 3 to 4.

\section{Krylov subspace methods for the overlap operator} \label{formsec}

\subsection{Notation and Basics}\label{BASICS}
The Wilson-Dirac fermion operator,
\begin{equation*}
M = I -\kappa D_W,
\end{equation*}
represents a nearest neighbour coupling on a four-dimensional
space-time lattice, where the `hopping term' $D_W$ is a
non-normal sparse matrix, see \eq{HOPPING} in the appendix. The
coupling parameter $\kappa$ is a real number which defines the
relative quark mass.

The massless overlap operator (using the Wilson operator as a kernel) is defined as
\begin{equation*} \label{overlap:def}
   D_0 = I + M \cdot (M^{\dagger}M)^{-\frac{1}{2}}.
\end{equation*}
For the massive overlap operator, for notational convenience, we use
a mass parameter $\rho > 1$ such that this operator is given as
\begin{equation} \label{overlapmass:def}
   D = \rho I + M \cdot (M^{\dagger}M)^{-\frac{1}{2}},
\end{equation}
with $\rho\ge 1$. How this form relates to Neuberger's choice and to the
quark mass is explained in the appendix, \eq{REGULAR}.

Expressing \eqnref{overlapmass:def} in terms of the hermitian Wilson fermion matrix
$Q = \gamma_5M$,
see \eq{HWD}, the overlap operator can equivalently be written as
\begin{equation*} \label{overlap2:def}
D = \rho I + \gamma_5 \sign{Q} = \gamma_5 \cdot (\rho \gamma_5 + \sign{Q}),
\end{equation*}
with $\gamma_5$ being defined in Appendix \ref{AppA} and $\sign{Q}$ being the
standard matrix sign function. Note that $\rho \gamma_5 + \sign{Q}$ is
hermitian, whereas $\gamma_5 \sign{Q}$ is unitary. To reflect these facts in our notation, we define
$$
D_u = \rho I + \gamma_5 \sign{Q},\qquad
D_h = \rho \gamma_5 + \sign{Q},
$$
where $D_u = \gamma_5 D_h$.

In a simulation with dynamical fermions, the costly computational task is
the inclusion of the fermionic part of the action into the `force' evolving
the gauge fields. This requires to solve linear systems of the form
\begin{equation} \label{squared_eq}
  D_u^{\dagger}D_u x = b \; \Longleftrightarrow \; D_h^2x = b.
\end{equation}
From a practical point of view this means that we want to find an approximate
solution $\hat x$ for \eqnref{squared_eq} such that
\begin{equation}
\label{main_eq}
\|D_h^2 \hat x - b\|_2 \le  {\mathcal O}(\epsilon).
\end{equation}
The value $\epsilon$ is prescribed and depends on the accuracy of
the overall process. In this paper we assume that this value is
given.

The major part of this paper is concerned with numerical methods
for the above `squared' equation, but we will occasionally also
consider the equation
\begin{equation} \label{propagator_eq}
  D_u x = b
\end{equation}
which has to be solved when computing propagators.

The standard solution method for solving the linear systems
(\ref{squared_eq}) or \eqnref{propagator_eq} is based on a nested
iteration scheme. The {\em outer} iteration consists of an
iterative linear system solver that invokes in every iteration
step a vector iteration method for approximating the action of the
matrix sign function to a vector. In the case of the squared
system \eqnref{squared_eq}, this {\em inner} iteration must even
be done twice.

\subsection{Adequate Krylov Methods}\label{sec:KrylovMethods}

In order to be self-contained, let us summarise results from
\cite{vdEetal:03a}, where Krylov subspace methods for the outer
iteration are discussed in detail.

Solving the propagator equation
\begin{equation} \label{prop1_eq}
D_u x = b
\end{equation}
is equivalent to solving the symmetrised equation
\begin{equation} \label{prop2_eq}
D_h x = \gamma_5 b = \hat{b}
\end{equation}
or one of the normal equations
\begin{equation} \label{prop3_eq}
D_h^2x = D_h \hat{b}, \enspace \mbox{ or } \enspace D^2_hy = \hat{b}, x = D_hy.
\end{equation}
Interestingly, for all these equations one has feasible optimal
Krylov subspace methods at hand, i.e.\ methods, which rely on short
recurrences and which obtain iterates satisfying an optimality
condition on the Krylov subspace generated by the matrix of the
respective equation: The normal equations \eqnref{prop3_eq} can be
solved with the CG method (its iterates minimise the error in the
energy norm), the symmetrised equation \eqnref{prop2_eq} can be
solved via the MINRES method (its iterates minimise the residual
in the 2-norm), and the shifted unitary system \eqnref{prop1_eq}
can be solved with a less well known method of Jagels and Reichel
\cite{JR94}, which we termed SUMR in \cite{vdEetal:03a} (its iterates
have again minimal residual in the 2-norm).

The theoretical results from \cite{vdEetal:03a}, backed up by
numerical experiments, show that solving \eqnref{prop1_eq} via
SUMR is the best of all these methods, resulting in savings of up
to 30\% as compared to the other two approaches which both require
approximately the same computational work.

When it comes to solving the squared equation
\begin{equation} \label{squared1_eq}
 D^2_hx = b \; \Longleftrightarrow  \; D_u^\dagger D_u x = b
\end{equation}
we have two basic options: Either solve \eqnref{squared1_eq} as it
stands, using the CG method for the hermitian and positive
definite matrix $D_h^2 = D_u^\dagger D_u$, or using a two pass
strategy solving
\[
D_h y = b, \; D_h x = y,
\]
or
\[
D^\dagger_u y = b, \; D_u x = y.
\]
From the previous discussion it is immediately clear that the
latter form of the two-pass strategy is to be preferred, and the
results from \cite{vdEetal:03a} further show that solving
\eqnref{squared1_eq} via CG is usually the best option.

\section{Strategies for the accuracy of the inner iteration} \label{inexactsec}

In the first paper of this sequence \cite{EFL02},
we discussed {\em a posteriori}
error estimators for various vector iteration methods that
construct approximations from a Krylov
subspace to the action of $\hbox{sign}(Q)$ to a vector.
This included Lanczos-type methods and computational
schemes based on the multi-shift CG method. The control over the
error of the matrix-vector products is very important in a
two-level iteration scheme and in this section we discuss how
to exploit this.
For generality, we consider the solution
of a generic linear system
\begin{equation*}
A x = b,
\end{equation*}
where $A$ and $b$ depend on the formulation used and $A$ involves somehow the matrix sign function of $Q$.
In step $j+1$ of the Krylov subspace
method we have to compute an approximation $\hat s$ to the product
of the matrix $A$ times a vector, say $y$, as
\begin{equation} \label{outer_accuracy:eq}
\|A y - \hat s\| \le \eta_j \cdot \|A\| \cdot \|y\|.
\end{equation}
An obvious choice is to pick $\eta_j$ fixed and equal to
$\epsilon$, the overall accuracy in \eqnref{main_eq}, in every iteration step.
 Since this can be seen as
raising the unit roundoff to a level of $\epsilon$, we expect that
(\ref{main_eq}) can be achieved in this case. However,
better strategies for choosing $\eta_j$ do exist.

In the past few years, various researchers have investigated the
effect of approximately computed matrix-vector products on Krylov subspace
methods. Outside the context of this paper, this plays a role in,
for example, electromagnetic applications \cite{Car02}, the
solution of Schur complement systems \cite{BFG00,SSz02a,ESG03}
and eigenvalue problems \cite{GZZ00}.  This work has led to,
so-called, `relaxation strategies' for choosing the $\eta_j$,
starting with the empirical results in \cite{BFr00a,BFG00} and
later followed by the more theoretical papers \cite{SEs02} and
\cite{SSz02a}.  The goal of these relaxation strategies is, given
a required residual precision of order $\epsilon$ (similar as in (\ref{main_eq})),
to minimise the total amount of work that is spent in the computation
of the matrix-vector products. It turns out that accurate
approximations to the matrix-vector product are required in the
very first iteration steps, but this precision can be relaxed as
the methods proceed (which explains the term
relaxation). In this section we summarise the main conclusions
which are of interest to nested iterations for the QCD overlap
formulation.

In a Krylov subspace method, for example the CG method,
in every iteration step an approximation to the residual, $r^k$,
and an iterate $x^k$ are computed, also when the matrix
vector product is not exact. Unfortunately, from the very first iteration on,
due to the approximate matrix-vector products, the true residual,
$b-Ax^k$, and the computed approximation to the residual, $r^k$,
drift apart. Therefore, the vector $r^k$ is not a good estimator
for the quality of the computed iterate.  The approach taken in
\cite{SEs02} is to consider the inequality
\[
\begin{array}{clclc}
\|\underbrace{b-Ax^k}\| &\le& \|\underbrace{r^k - (b-Ax^k)}\| &+&
\|\underbrace{r^k}\|.\\
\hbox{true residual} & & \hbox{residual gap} & & \hbox{computed residual}
\end{array}
\]
The computed residual norms $\|r^k\|$ can be monitored during the
iteration process and there is overwhelming numerical -- and
partially theoretical -- evidence that the computed residuals  initially decrease
and stagnate in the end at a level smaller than the size of the
unknown residual gap. From a practical point of view, this means
that strategies for controlling the error of the matrix sign
function can be derived by bounding the size of the gap in terms
of the $\eta_j$ and subsequently choosing the $\eta_j$ such that
the size of the residual gap does not become larger than the order
of $\epsilon$. This approach is taken in \cite{SEs02,ESG03} and
it confirms and leads to improvements upon the empirically found strategies proposed by
Bouras et al.\ \cite{BFr00a,BFG00}.
For clarity we discuss this in more detail for the CG method where
the matrix $A$ is hermitian positive definite.

In the CG method the iterate and residual are updated using the formula
\begin{equation*}
r^j = r^{j-1} - \alpha_{j-1} q^{j-1}, \quad x^j = x^{j-1} + \alpha_{j-1} p^{j-1}
\end{equation*}
with
\begin{equation*}
\|q^{j-1} - A p^{j-1} \| \le \eta_{j-1} \cdot \|A\| \cdot  \|p^{j-1}\| \quad \hbox{and} \quad
\alpha_{j-1} = \frac{\|r^{j-1}\|^2}{{q^{j-1}}^\dagger \cdot p^{j-1}}.
\end{equation*}
A simple inductive argument shows that
\begin{equation*}
\|r^k - (b - Ax^k)\| \le   \sum_{j=0}^{k-1} \eta_j |\alpha_j| \cdot  \|A\| \cdot \|p^j\|.
\end{equation*}
To continue we need to bound $|\alpha_j| \cdot \|p^j\|$ and
to keep our discussion pertinent we will start by considering the size of these quantities in case
of exact matrix-vector multiplications.
From the definition of $\alpha_j$, we have
\begin{equation}\label{afschatteneq}
|\alpha_j| \cdot  \|p^j\| = \frac{\|p^j\|^2}{|{p^j}^\dagger\cdot  q^j|} \cdot \frac{\|r^j\|^2}{\|p^j\|}.
\end{equation}
It is straightforward to bound  the first term in (\ref{afschatteneq}). Using $q^j = Ap^j$, we see that with an exact multiplication
this term is smaller than $\|A^{-1}\|$.
Furthermore, using the recursion of the conjugate search directions
\begin{equation*}
p^j = r^{j} -  \gamma_j/\gamma_{j-1} p^{j-1}\quad \hbox{and} \quad
\gamma_{j} = \|r^{j}\|^2,
\end{equation*}
it follows by exploiting orthogonality properties that
\begin{equation*}
\|p^j\| = \|r^j\|^2 \cdot \sqrt{\sum_{i=0}^j \|r^i\|^{-2}}.
\end{equation*}

As is explained in \cite{SEs02} (giving the details would be beyond the scope of this paper),
it is reasonable to assume that in the case of an inexact matrix-vector product
there is a modest constant $c$ such that right hand side of \eqnref{afschatteneq} times $c$
is an upper bound for $|\alpha_j| \cdot \|p^j\|$.

With this assumption we have that the residual gap after $k$ steps is bounded as
\begin{equation*}
\|r^k - (b - Ax^k)\| \le  c \cdot\|A\| \cdot \|A^{-1}\| \sum_{j=0}^{k-1} \eta_j \rho_j, \quad \rho_j = (\sum_{i=0}^j \|r^i\|^{-2})^{-1/2}.
\end{equation*}
Since, we are interested in a final residual precision of about $\epsilon$ our strategy is to keep the residual gap of this size. Hence,
we propose, following \cite{SEs02,ESG03} which improved upon the empirical strategy from \cite{BFG00},
\begin{equation}\label{relaxcgeq}
\eta_j = \frac{\epsilon}{\rho_j},
\end{equation}
which guarantees that
\begin{equation}\label{resgapcgeq}
\|r^k - (b - Ax^k)\| \le  c \cdot k\cdot\|A\| \|A^{-1}\| \epsilon.
\end{equation}
For the purpose of illustration, Figure~\ref{rel_cg_alg} gives an algorithmic description
of the CG method with this strategy for tuning the errors in the matrix-vector products.\footnote{Matlab code for all methods presented in this paper
is publicly available through the Internet at {\tt www.uni-wuppertal.de/org/SciComp/preprints}}

\begin{figure}
\centerline{
\begin{minipage}{\linewidth}
\hrulefill
\begin{alg} $\hbox{\rm RelCG}(A,b,\epsilon)$
\begin{algorithmic}
\STATE \COMMENT{\textup{computes} $x$ \textup{with} $\|Ax - b\| \le \epsilon\cdot \|b\|$ \textup{via relaxed CG}}
\STATE
\STATE $x=0;$ \qquad \{\textup{initial value}\}
\STATE $r=b;$
\STATE $p=r;$
\STATE $\gamma_{old}=\gamma=r^\dagger\cdot r;$
\STATE $\zeta=1/\gamma;$
\WHILE{$\sqrt{\gamma} > \epsilon\cdot \|b\|$}
\STATE \textup{compute} $q$ \textup{with} $\|A p - q\| \le \epsilon\cdot \|b\|\cdot \|p\|\cdot \sqrt{\zeta};$
\STATE $\beta=q^\dagger\cdot p;$
\STATE $\alpha=\gamma/\beta;$
\STATE $x=x + \alpha\cdot p;$
\STATE $r=r - \alpha\cdot q;$
\STATE $\gamma=r^\dagger\cdot r;$
\STATE $\zeta=\zeta+1/\gamma;$
\STATE $p=r + \gamma/\gamma_{old}\cdot p;$
\STATE $\gamma_{old}=\gamma;$
\ENDWHILE
\end{algorithmic}
\end{alg}
\hrulefill
\end{minipage}
}
\caption{generic relaxed CG \label{rel_cg_alg}}
\end{figure}

If we assume that the computed residuals $r^j$ decrease and we terminate in step $k$ where $\|r^k\|$ is smaller than $\epsilon$ then
the size of the true residual $\|b-Ax^k\|$ is bounded by (\ref{resgapcgeq}) plus $\epsilon$. This shows that we have achieved
our accuracy goal despite the fact that we work with less accurate matrix-vector products as the iterative process proceeds.
Notice that we have not given {\em a priori} guarantees that the computed residuals become smaller than $\epsilon$
(with a comparable speed to the exact process) and, furthermore, that $c$ is not a very large constant.
Unfortunately, we are not aware of the existence of rigorous estimates for these quantities in the context of relaxation
(there are results for the case that $\eta_j$ is constant, see \cite{Gre97}). However, numerous numerical experiments show that this is not an
issue in practice and the advantage of this way of deriving strategies for picking $\eta_j$ is that these quantities
can be monitored at little additional cost, if necessary.

One issue remains if we want to apply the discussed relaxation strategy for solving \eqnref{squared1_eq}.
In this case we must be able to assess the accuracy of our computed approximation $\hat{s}$
to $D_h^2 y$ through the accuracy in computing the action of
$\sign{Q}$ on a vector, see \eqnref{outer_accuracy:eq}. We do so by expanding $D_h^2$ as
\[
D_h^2 = (\rho^2+1)I + \rho \gamma_5 \sign{Q} + \rho \sign{Q}\gamma_5,
\]
so that we achieve $\| D_h^2y - \hat{s}\| \leq \eta \|y\|$ by
requiring that the approximations $\hat{s}_1$ to $\sign{Q}y$
and $\hat{s}_2$ to $\sign{Q} (\gamma_5 y)$ fulfil
\[
\| \sign{Q}y - \hat{s}_1\| \leq \frac{1}{2\rho} \eta \|y\|
   \enspace \mbox{ and }
   \| \sign{Q}(\gamma_5y) - \hat{s}_2\| \leq \frac{1}{2\rho} \eta \|y\|.
\]

\subsection{Relaxation strategies for SUMR and MINRES}

So far, we have discussed relaxation for the CG method since this is fairly straightforward due to its two-term recurrences. In \cite{SEs02} a general framework is given that allows an analysis for a large variety of
Krylov subspace methods. We refer the reader to this paper for more information. In general, the relaxation strategies
proposed there for these methods guarantee bounds on the residual gap of the form (\ref{resgapcgeq}).
In Table~\ref{method_table} we have summarised the strategies for
choosing the $\eta_j$ for the Krylov subspace methods relevant
for the formulations in the previous section.

\begin{table}
\begin{small}
\begin{center}
\begin{tabular}{l|lll}
\hline
matrix properties & method & tolerance $\eta_j$ & reference\\
\hline
herm.\ pos.\ def.\ & CG & $\eta_j = \epsilon \sqrt{\sum_{i=0}^j \|r^i\|^{-2}}$ & equation~\eqnref{relaxcgeq}\\
$(D_h^2$, equation \eqnref{squared1_eq}) &  & & \\
herm.\ indefinite & MINRES & $\eta_j = \epsilon/\|r^j\|$ & \cite[p. 20]{SEs02}\\
$(D_h$, equation \eqnref{prop2_eq}) &  & & \\
shifted unitary & SUMR & $\eta_j = \epsilon/\|r^j\|$ \\
$(D_u$, equation \eqnref{prop1_eq}) &  & & \\
\hline
\end{tabular}
\vspace*{.3cm}
\caption{Advised Krylov subspace method and corresponding strategy for
  tuning the precision of the matrix-vector products as a function
  of the properties of the matrix $A$.}
\label{method_table} \end{center}
\end{small}
\end{table}

For the SUMR method mentioned in the previous section, no relaxation strategies
have been proposed so far. Unfortunately, an analysis of SUMR with approximate matrix-vector products is more involved
than for the other iterative solvers
since no residuals are computed during the iterative process (only
their lengths are available). However, for theoretical purposes we can
introduce an additional recursion for the residual vectors  and consider the residual gap.
It is then possible to show that the same results apply as for the full GMRES method
in \cite[Section~7]{SEs02}. Without going into details, let us just state that therefore
we expect good results for SUMR using the
same strategy as used for the GMRES method, see Table~\ref{method_table}.

\begin{figure}
\centerline{\includegraphics[scale=0.4]{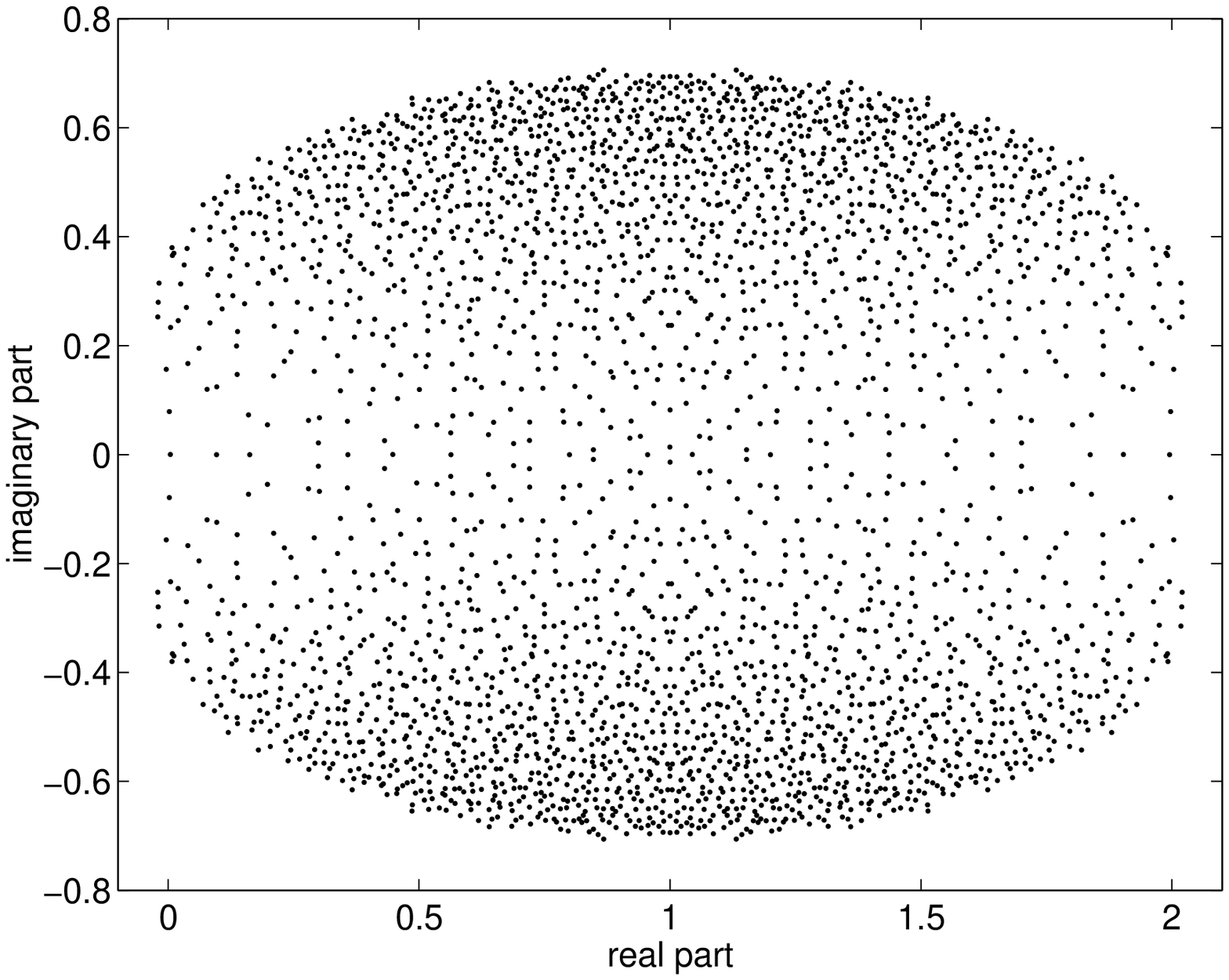}
 \includegraphics[scale=0.4]{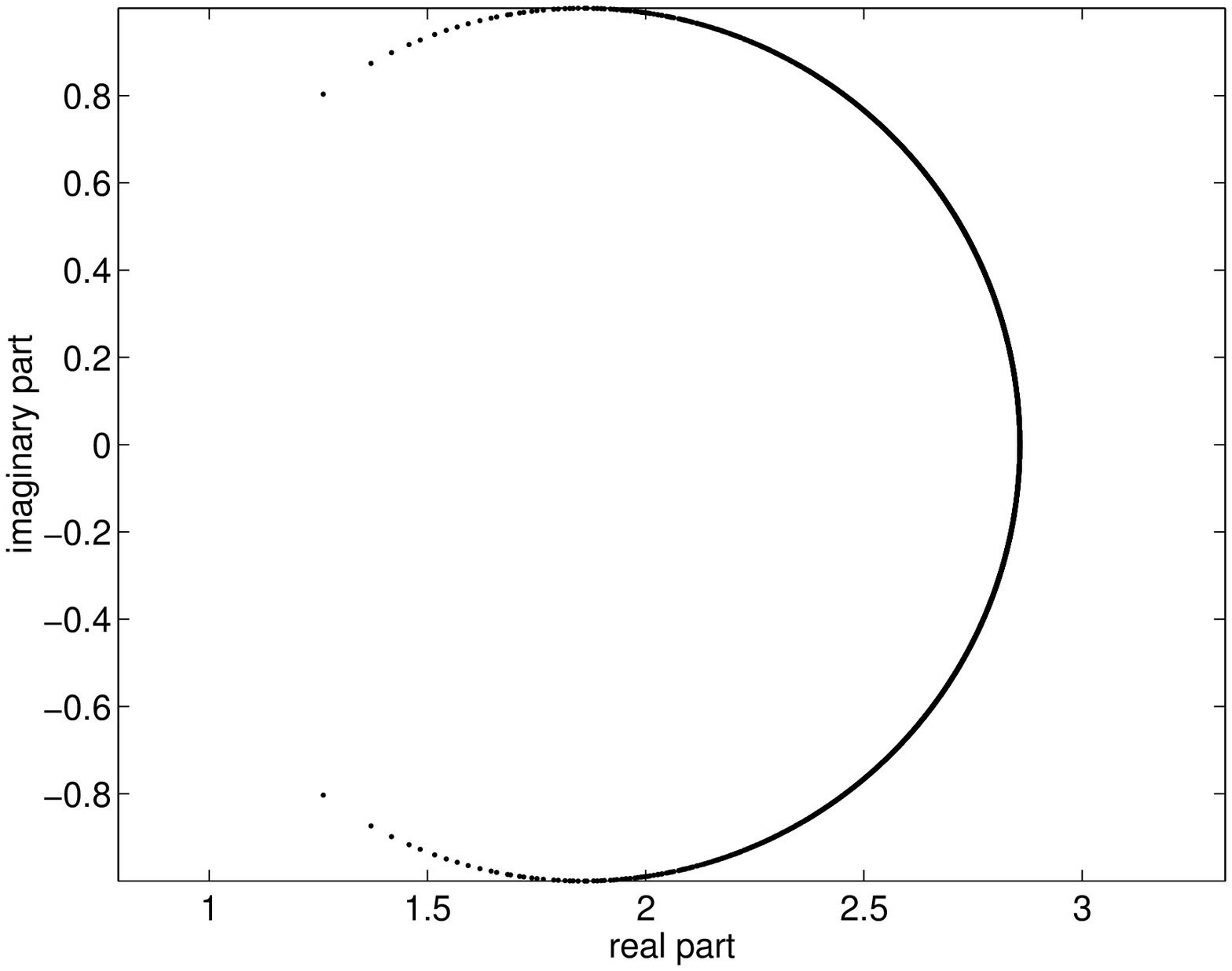}
}
\caption{Spectrum of the Wilson fermion matrix $M$ for our $4^4$ configuration (left),
spectrum of $D_u$ for $\mu = 0.3$ (right).
 \label{Wilson_fig}}
\end{figure}

\begin{figure}
\centerline{\includegraphics[scale=0.4]{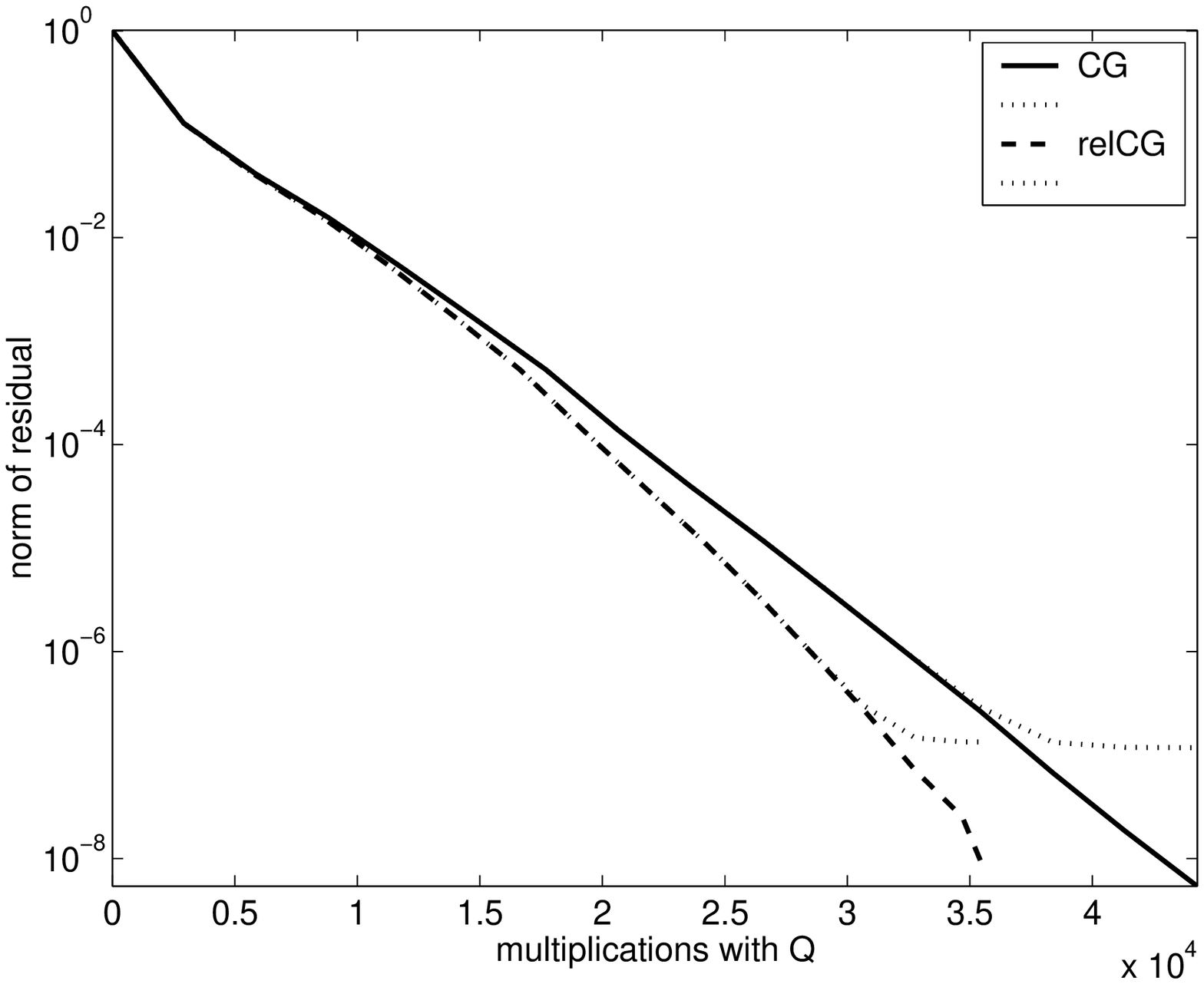}
\hfill \includegraphics[scale=0.4]{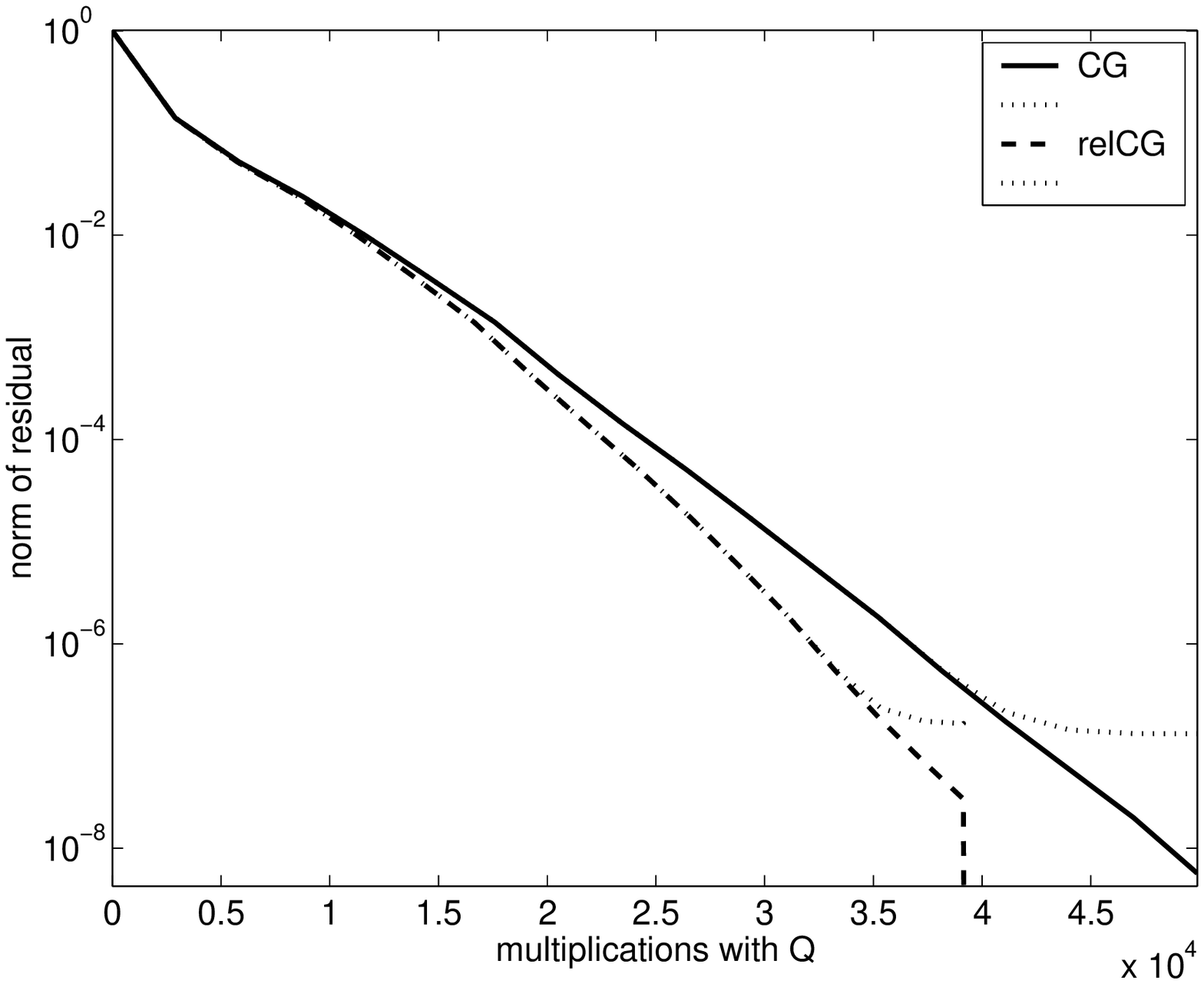}
}
\caption{Relaxed and non-relaxed CG for \eqnref{squared1_eq}. Left: $\mu = 0.3$. Right: $\mu = 0.1$.
\label{relnonrel_fig}
}
\end{figure}

\subsection{Numerical illustration}
To illustrate the effects of relaxation, we here report on
numerical experiments for a simple test situation. More extensive experiments will be
reported in Section~\ref{numericalsec}. We use the $4^4$ example configuration Conf1
from Section~\ref{numericalsec}, which is the result of a dynamical simulation at
$\beta = 5.4$. The hopping parameter in the Wilson fermion matrix was taken as
$\kappa = 0.2$.
The spectrum of the Wilson fermion matrix $M$ is given in the left part of Figure~\ref{Wilson_fig},
the spectrum of the overlap operator $D_u$ is plotted on the right.
We solved the `squared' equation \eqnref{squared1_eq} using the CG method for
the mass parameters $\mu = 0.3$ and $\mu = 0.1$, where $\rho = (1+\mu)/(1-\mu)$, see \eq{REGULAR}.

The plots in Figure~\ref{relnonrel_fig}
give the norm of the (computed) residual
as a function of the number of
matrix-vector multiplies (MVMs) with $Q$.
These MVMs all occur in the multi-shift CG method when approximating $\sign{Q}x$ via the
Zolotarev approach. Details of our implementation are given in Section~\ref{numericalsec}. Each plot contains two convergence curves, one for CG without relaxation, i.e.,
with a fixed precision for the MVMs,
and one with the relaxed CG method described in Algorithm~\ref{rel_cg_alg}.

In the relaxed CG methods we also used a high accuracy
inner iteration for computing the sign function to compare the true and the computed residuals.
The true residuals are plotted in Figure~\ref{relnonrel_fig} as dotted lines. We see that the true and the computed residuals are
virtually the same until they are down to $\epsilon = 10^{-6}$, the required accuracy, which was the parameter used in \eqnref{relaxcgeq}.
We see that the relaxation strategies
yield an improvement in the order of 20\%, regardless of the value of $\mu$.
This improvement is larger in the more realistic computations to be reported in Section~\ref{numericalsec}. There we perform an additional eigenvalue projection step to speed up the overall computation. In this situation relaxation leads to larger gains ranging from 30\% to 40\%.

\section{Further improvements: Recursive preconditioning} \label{precsec}
In the previous section we discussed strategies for controlling
the error of the matrix-vector products. The preliminary numerical experiments
there showed that a reduction of at least 20\% can be expected compared
to the case of using a fixed precision for the matrix-vector
products in all steps.  Two important practical observations
should be made, see also \cite[Section~3]{ESG03}.  First, we note
that, if the number of iterations to reach the desired residual
reduction is large, then there can be a considerable accumulation
of the errors in the matrix-vector product in the residual gap.
This is reflected for the CG method  by the
fact that the required number of iterations appears in the upper
bound on the residual gap in \eqnref{resgapcgeq}  as a factor $k$.  In
practice, this might mean that the tolerance on the matrix-vectors
has to be decreased, which is the tactic taken in \cite{SSz02a}.
But more importantly, the strategies discussed in the
previous section, take the error in the matrix-vector products,
essentially, inversely proportional to $\|r^j\|$.  Krylov subspace
methods often show {\em superlinear} convergence, meaning that the
convergence speed increases as the iterative method proceeds.
Hence, the number of inexpensive approximations to the
matrix-vector products is relatively small and this limits the
maximal gain that can be achieved with the relaxation approach.

Stating the same observation from a different point of view, we
conclude that the relaxation strategy should work particularly
well when the convergence of the iteration is fast (but linear) from the very
beginning.  In order to achieve this, we investigate the idea of
`preconditioning' the Krylov subspace method by another (inexact)
Krylov subspace method set to a larger tolerance of $\xi_j$ in
step $j+1$. We refer to this as {\em recursive preconditioning}.
\begin{figure}
\centering
\includegraphics[scale=0.8]{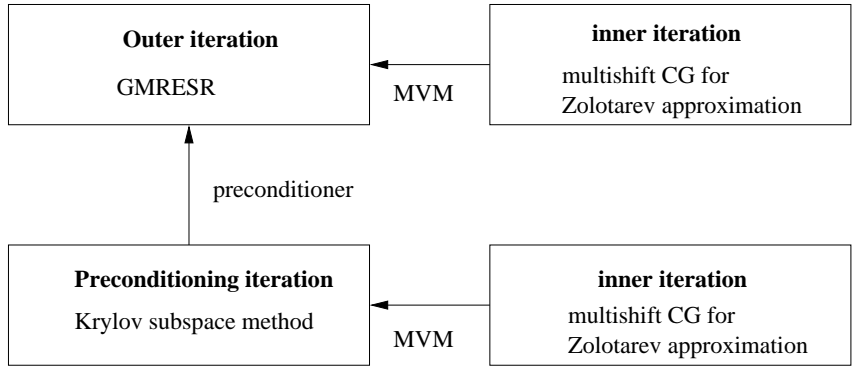}
\caption{Overview of our recursive preconditioning computational scheme
\label{schemafig}}
\end{figure}
To stay consistent with the terminology used so far, we refer to
the inexact Krylov method and its variable preconditioner as the
{\em outer iteration} and {\em preconditioning iteration}
respectively, reserving the term {\em inner iteration} to the
method which approximates the matrix vector product. The inner
iteration is thus used in both, the preconditioning and the outer
iteration, see Figure~\ref{schemafig}.

Methods that can be used for the outer iteration are the so-called
{\em flexible methods}. These are methods that are specially
designed for dealing with variable preconditioning, e.g.,
\cite{GOv88,Saa93,VVu94}. In \cite{ESG03} these methods were
combined with approximate matrix-vector products.  For our
numerical experiments we chose the GMRESR (GMRES recursive, \cite{VVu94})
method as the outer iteration. Note that
other choices like
flexible GMRES \cite{Saa93} are an equally good option.
We have chosen GMRESR here since it is slightly more straightforward to implement.
The paper
\cite{ESG03} analyses various choices for the accuracies $\eta_j$
and $\xi_j$ and shows that $\eta_j = \epsilon/\|r^j\|$ and $\xi_j
= \xi$ fixed are good choices.  Figure~\ref{rec_alg_fig} gives a
Matlab-style algorithmic description of the overall method, which we call {\em
relaxed GMRESR}. To stress the preconditioning iteration, we sometimes
  add it in parenthesis, so that, e.g., {\em relaxed GMRESR(CG)} means that we use
  the CG method as the preconditioning iteration.

\begin{figure}
\begin{minipage}{\linewidth}
\hrulefill
\begin{alg} $\hbox{\rm relGMRESR}(A,b,\epsilon)$
\begin{algorithmic}
\STATE \COMMENT{\textup{computes} $x$ \textup{with} $\|Ax - b\| \le \epsilon\cdot \|b\|$ \textup{via relaxed  GMRESR}}
\STATE
\STATE $x=0;$ \qquad \{\textup{initial value}\}
\STATE $r=b;$
\STATE $C=[];$ \qquad \{\textup{empty matrix}\}
\STATE $U=[];$ \qquad \{\textup{empty matrix}\}
\WHILE{$\|r\| > \epsilon\cdot \|b\|$}
\STATE solve $Au = r$ to relative accuracy $\xi$ (for example $u=\hbox{\rm relCG}(A,r,\xi);$)
\STATE \qquad \qquad \qquad \{\textup{preconditioner}\}
\STATE \textup{compute} $c$ \textup{with} $\|A u - c\| \le \epsilon\cdot\|b\|\cdot \|u\| / \|r\|;$
\FOR{i=1:size(C,2)}
\STATE $\beta=C[:,i]^\dagger\cdot c;$
\STATE $c=c - \beta\cdot C[:,i];$
\STATE $u=u - \beta\cdot U[:,i];$
\ENDFOR
\STATE $c=c/\|c\|;$
\STATE $u=u/\|c\|;$
\STATE $C=[C,c];$
\STATE $U=[U,u];$
\STATE $\alpha=c^\dagger\cdot r;$
\STATE $x=x + \alpha\cdot u;$
\STATE $r=r - \alpha\cdot c;$
\ENDWHILE
\end{algorithmic}
\end{alg}
\hrulefill
\end{minipage}
\caption{Relaxed GMRESR \label{rec_alg_fig}}
\end{figure}

The recursive application of
iterative solution methods is often encountered
in scientific computing applications.
For example, van der Vorst and Vuik notice in \cite{VVu94} that
preconditioning GMRES with a fixed number of iterations of GMRES
can give a considerable improvement over restarted GMRES.
This explains the name `GMRESR' which we keep in this paper although
we use other choices for the preconditioner.

In the
context of approximate matrix-vector products, nested iterations
have been used by Carpentieri in his PhD thesis \cite{Car02}. He
uses flexible GMRES in the outer iteration and GMRES in the inner
iteration for an application from electromagnetics where the
matrix-vector products are approximated using a fast multipole
technique set to a fixed precision.  The paper \cite{ESG03} shows
numerical experiments for a Schur complement system that stems
from a model of global ocean circulation. Using the relaxed
preconditioned approach one gets a significant reduction in the
amount of work spent in the matrix-vector products.

A related idea for the QCD overlap formulation has recently been
advanced by Giusti et al. \cite[Section~9]{GHL02} in a method
which they call an {\em adapted-precision inversion algorithm}.
Their scheme corresponds to the approach presented here if,
instead of GMRESR, one
takes a simple Richardson iteration as the outer iteration and if,
in addition, the residuals are computed directly. The authors of
\cite{GHL02} do not discuss specific choices for the precision of
the matrix-vector product in the outer iteration and use a fixed
precision in the preconditioning iteration.  Our more general
approach allows the use of more sophisticated outer iterations
like GMRESR and, moreover, gives a specific and computationally
feasible strategy on how to choose the precision of the inner
iteration.

It is also interesting to mention that the idea of adapted
precision inversion is related to an earlier approach of
Hern\'andez et al. in \cite{HJL00} and Bori\c{c}i in \cite{Bor99}.

\section{Numerical experiments} \label{numericalsec}

Next we present numerical experiments carried out in a realistic setting.
To this purpose we have developed a Hybrid Monte-Carlo program (HMC) with
either one or two
flavours of dynamical overlap fermions based on the Zolotarev partial
fraction expansion~\cite{EFL02}.  Details as to the
construction of the overlap fermion force within the HMC can be found in
Appendix \ref{APPENDIXHMC}.

We have generated decorrelated configurations with one flavour
(see Section~\ref{chiralproj_sec} below) of dynamical
overlap fermions on an $8^4$-lattice at $\beta = 5.6$, and with two
flavours on a $4^4$
lattice at $\beta = 5.4$.  For our
experiments the Wilson kernel mass parameter has been adjusted to $\kappa =
0.2$. It is known that the
locality properties and spectral density of the overlap operator
depend strongly on the value of $\kappa$ used, with $\kappa \sim 0.2$
being the optimum value, at least in the quenched
theory on large lattices~\cite{Hernandez:1998et}.

We have used a mass parameter $\mu$~\cite{hep-lat/9710089}, which,
according to \eq{REGULAR}, is equivalent to $\rho
= (1+\mu)/(1-\mu)$, and we have chosen $\mu = 0.1$ ($\rho =
1.22$), and $\mu = 0.3$ ($\rho = 1.857$).  These values of
$\mu$ are similar to the smallest non-zero eigenvalues of the overlap
operator and, given our small lattices,  there will be little change in the
results when moving $\mu$ to smaller valence mass.

Our results will be given
for five configurations for the $4^4$ lattice volumes, separated by 50
HMC sweeps, and on  three plus  five configurations, separated by 20
HMC sweeps on the $8^4$ lattice. Additionally, some computations were performed on a
configuration from a quenched ($\beta = 6.0$) ensemble, with the
inversions performed at $\kappa = 0.2$, with three values of $\mu$, $0.3$, $0.1$
and $0.03$.

All our computations were carried out on the Wuppertal cluster computer ALiCE, using
16 processors for the calculations on the $8^4$ and $16^4$ lattices, and one processor
for the $4^4$ calculations.

\subsection{Projecting out low-lying eigenvectors}\label{sec:5.1}
Let $a$ and $b$ denote the smallest and largest eigenvalue of $Q^2$, respectively.
Then, in principle, we need the Zolotarev approximation to the sign function for
the domain $[-\sqrt{b},-\sqrt{a}] \cup [\sqrt{a},\sqrt{b}]$. As $a$ gets smaller,
we need an increasing number of terms in the Zolotarev approximation in order to obtain a
given accuracy. It is possible~\cite{xxxx} to
accelerate the calculation of the sign function by calculating the
$n_p$ `smallest' eigenvectors of the Wilson operator, and by treating
them exactly.  The Zolotarev approximation is then only needed on a domain
$[-\sqrt{b},-\sqrt{a'}] \cup [\sqrt{a'},\sqrt{b}]$ with $a'$ being not larger than the
$n_p+1^{\rm st}$ smallest eigenvalue of $Q^2$.

Besides from allowing us to shrink
the domain over which we need an accurate approximation
to the sign function, projecting out small eigenvectors also improves the condition number of
the Wilson operator $Q$. As a consequence the multi-mass inversion to be performed for the Zolotarev approximation converges
faster.

There are two different ways in which we can project out the
eigenvalues -- either out of the sign function, or out of the multi-mass solver
for the partial fraction expansion. Our preference is to fix $a'$ at
some suitable value, and to project the eigenvalues below $\sqrt{a'}$
directly out of the sign function, and those eigenvalues above $\sqrt{a'}$ out of
the multi-mass solver~\cite{paper4} (see Appendix \ref{APPEIG}).

We calculate the coefficients of the Zolotarev expansion so that the
sign function is approximated to machine precision within the range
$[\sqrt{a'},\sqrt{b}]$ (see Appendix~\ref{APPENDIXHMC}).
 We cannot vary the values of $a'$ and $b$ in the outer iteration
across the trajectory of the HMC algorithm without violating detailed
balance. Some fine tuning of the optimal
value of $a'$ is possible, but this optimisation lies outside the scope
of this paper: here we keep $b$ fixed at $10$, and $a'$ fixed at
$10^{-5}$. For all the calculations described in this section, we took $n_p =
28$, which always resulted in a $n_p+1^{\rm st}$ smallest eigenvalue larger than $a' = 10^{-5}$,
as required. The performance gain obtained from
the eigenvalue projection is briefly described in
Appendix \ref{APPEIG}.

In the preconditioner, $b$ and $a'$ were
allowed to vary. We took $b$ as the largest eigenvalue of $Q^2$ (usually
around 5), while $a'$ was the $n_p+1^{\rm st}$ smallest eigenvalue (around $10^{-3}$ for the $4^4$ lattices, and $10^{-4}$ for
the $8^4$ lattices). Since we need the sign function less precise in the preconditioner,
the number of poles in the Zolotarev approximation could be taken
quite small (see Appendix \ref{APPENDIXHMC}),
thus minimising the computational effort in the preconditioner.

\subsection{Results for the squared overlap operator}

\begin{figure}[t]
\hfill \includegraphics[width =6.5cm,height =
  5cm]{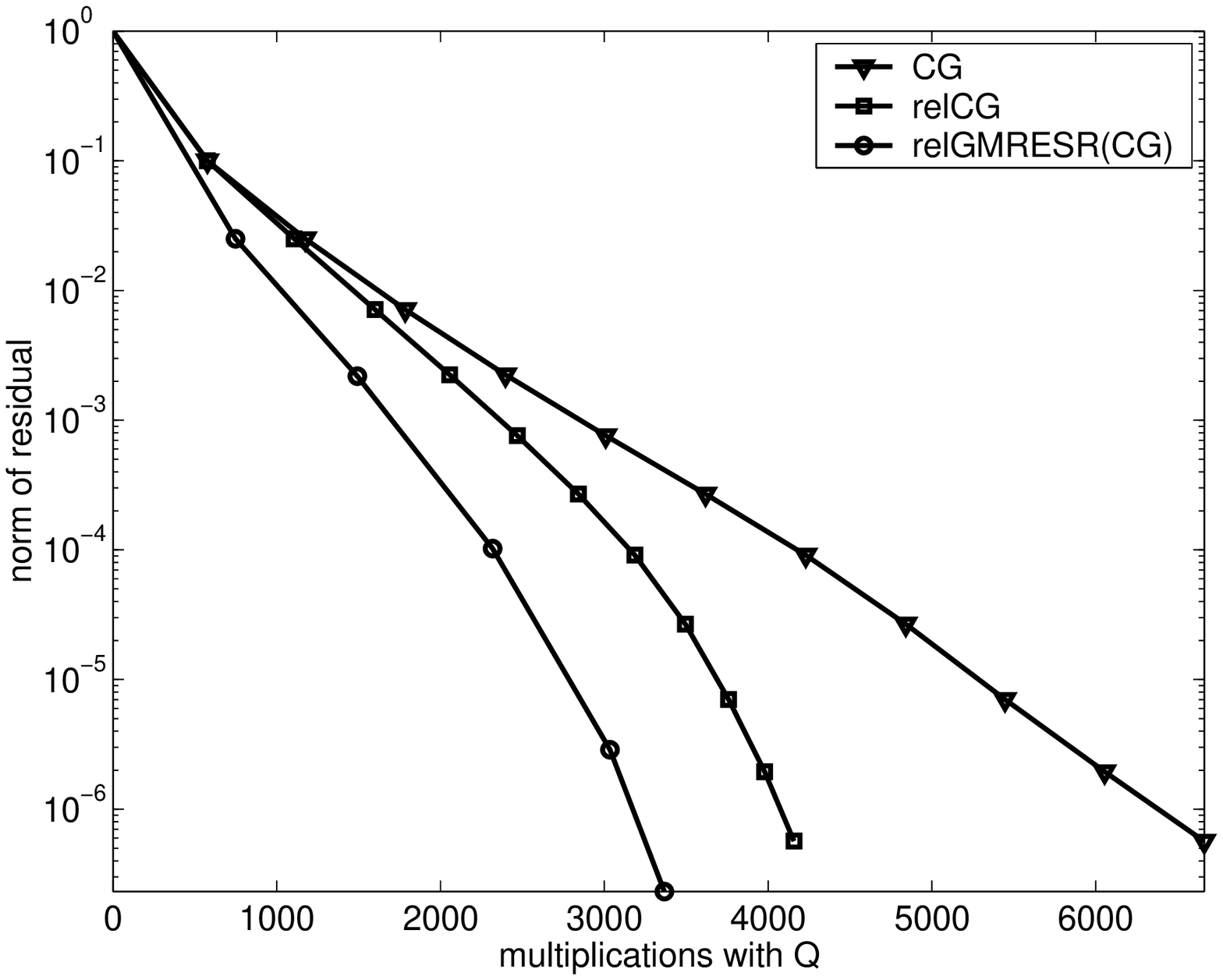}
 \mbox{} \hfill
\includegraphics[width =6.5cm,height =
  5cm]{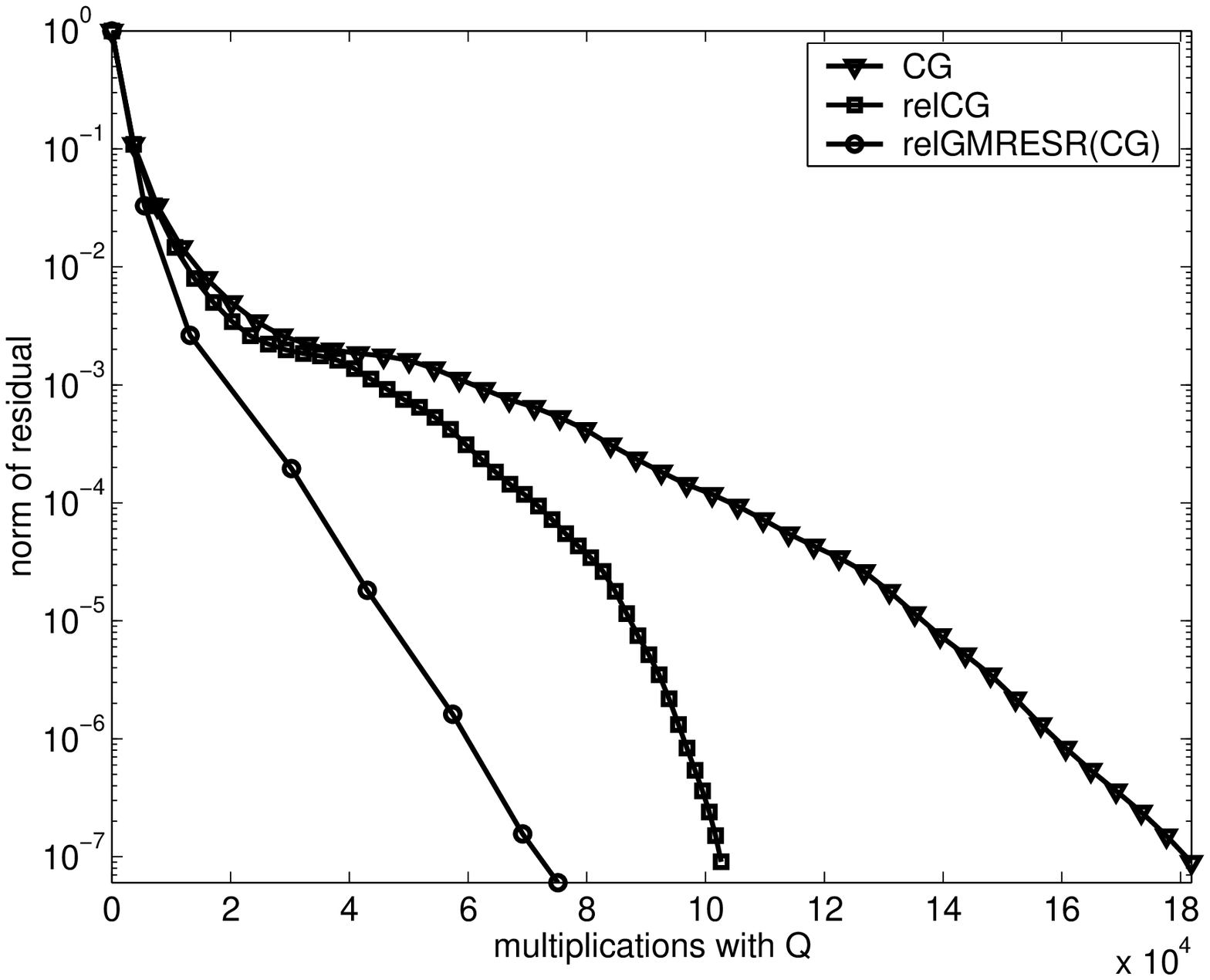} \hfill \mbox{} \vspace*{0.2cm}
  \\
  \mbox{}\hfill \begin{minipage}{4.5cm}First $4^4$, $\mu = 0.3$ configuration from table~\ref{tab:times1b}
      \end{minipage}
  \hfill \begin{minipage}{4.5cm}
         Third $8^4$, $\mu = 0.1$ configuration from table~\ref{tab:times2}
		 \end{minipage}
  \hfill \mbox{}\\
\caption{Convergence history for unrelaxed CG, relaxed CG and
  relaxed GMRESR for one inversion of the squared equation
  \eqnref{squared_eq}.
  We plot the norm of the residual vs.\ the number of
  calls to the Wilson operator $Q$. The tics indicate each (outer) iteration:
  On the $4^4$ lattice relaxed GMRESR needs 5
  iterations to converge, whereas unrelaxed and relaxed CG both require 11 iterations.}
  \label{fig:cg1} \label{fig:cg2}
  \vspace*{0.3cm}
\end{figure}

Figure~\ref{fig:cg1} gives the convergence history for the third
configuration on the $8^4$ and the first on the $4^4$ lattice. The plots
show the residual norm against the numbers of MVMs with the Wilson
fermion matrix $Q$. Since the latter dominate the performance of the entire process, we can
consider them as a first approximation of the overall execution time.
The figure compares the unrelaxed CG, relaxed CG  and the
relaxed GMRESR(CG)
methods. In addition to the plots, we give the actual timings for our implementations
in Tables \ref{tab:times1}-\ref{tab:times2b}. Note that for the preconditioned iterations
the gain in time is more than the gain obtained in MVMs with $Q$. The reason is that in
the preconditioner we only have five different shifts to work on when performing
the multi-mass inversion for the Zolotarev approximation, as opposed to the 25 shifts to be used
in the outer iteration.

Comparing the times from the tables we see that on
the $8^4$ lattices relaxation already reduces the computational effort
by a factor of 1.7. Additional preconditioning
further reduced the effort by an additional factor of 2.2
(taking $\xi_j=\xi = 0.1$
independently
of $j$ since there was little change in the gain for $0.01<\xi_j < 0.3$).

The gain turned out to be smaller
on the $4^4$ ensembles, with  only about a factor of 1.5 gain for the
relaxation, and an additional factor of 1.3 for the preconditioner.
The gain on the $16^4$ quenched configuration is similar to the gain
for the $8^4$ ensembles. The lower gain for the preconditioner on the smaller lattices is due to the fact that the CG inversion on $4^4$ lattices
converges already quite fast, a consequence of the particular eigenvalue distribution of the $4^4$ lattice already observed in Figure~\ref{Wilson_fig}. Thus, the $4^4$ inversion spent more
time in the outer GMRESR algorithm (compared to the time spent in the
CG preconditioner) than the $8^4$ inversion did. Since one
sweep through the GMRESR algorithm takes considerably longer than one
sweep through the CG preconditioner, this reduced the gain of the
preconditioning on the smallest lattices. For the same reason, the
gain achieved by
preconditioning is larger as we decrease the overlap mass, especially
on the larger lattices, where the overlap operator is less well
conditioned. For example, on the $16^4$ quenched configuration, the
gain for using the preconditioner is 1.75 times larger for $\mu =
0.03$ than for $\mu = 0.3$.

\begin{table}
\begin{small}
\begin{center}
\begin{tabular}{l l l l l l}
Method&Conf 1&Conf 2&Conf 3&Conf 4&Conf 5\\ \hline
CG            &53&55&53&57&55
\\
relCG         &34(1.56)&35(1.57)&36(1.47)&37(1.54)&38(1.45)
\\
relGMRESR(CG) &19(2.78)&20(2.75)&24(2.21)&25(2.28)&26(2.11)
\\
\hline
\end{tabular}
\\
$\mu = 0.3$ \\
\bigskip

\begin{tabular}{l l l l l l}
Method&Conf 1&Conf 2&Conf 3&Conf 4&Conf 5\\
\hline
CG            &50&46&44&46&48
\\
relCG         &33(1.52)&31(1.48)&30(1.47)&32(1.44)&31(1.55)
\\
relGMRESR(CG) &22(2.27)&20(2.30)&23(1.91)&25(1.84)&21(2.29)
\\
\hline
\end{tabular}
\\
$\mu=0.1$
\end{center}
\vspace{.2cm}
\caption{Times (in seconds) for one inversion on the five $4^4$
configurations with  $\beta = 5.4$, run on 1 processor of ALiCE. The number in brackets is
 the gain from the unrelaxed and unpreconditioned (CG) inversion.
 \label{tab:times1b}
 \label{tab:times1}
 }
\vspace{.2cm}
\end{small}
\end{table}

\begin{table}
\begin{small}
\begin{center}
\begin{tabular}{l l l l l l}
Method&Conf 1&Conf 2&Conf 3&Conf 4&Conf 5\\
\hline
CG            &1419&1139&1216&1307&1305
\\
relCG         &754(1.88)&697(1.63)&737(1.65)&816(1.60)&767(1.70)
\\
relGMRESR(CG) &319(4.45)&301(3.78)&315(3.86)&364(3.59)&341(3.82)
\\
\hline
\end{tabular}
\\
$\mu = 0.3$
\bigskip

\begin{tabular}{l l l l}
Method&Conf 1&Conf 2&Conf 3\\
\hline
CG            &1965&2052&2039
\\
relCG         &1202(1.63)&1250(1.64)&1234(1.65)
\\
relGMRESR(CG) &614(3.20)&567(3.61)&547(3.72)
\\
\hline
\end{tabular}
\\
$\mu = 0.1$
\end{center}
\vspace*{.2cm}
\caption{Times (in seconds) for one
inversion on the three $8^4$
configurations with $\beta = 5.6$, run on 16 processors of ALiCE.
\label{tab:times2}
\label{tab:times2b}}
\vspace{.2cm}
\end{small}
\end{table}

\begin{table}
\begin{small}
\begin{center}
\begin{tabular}{l l l l}
Method&$\mu = 0.03$&$\mu = 0.1$ & $\mu = 0.3$
\\
\hline
CG                &31430&9022&3493
\\
relCG                  &18813(1.67)&5981(1.51)&2610(1.34)
\\
relGMRESR(CG)  &6642(4.73)&2329(3.87)&1286(2.71)
\\
\hline
\end{tabular}
\end{center}
\vspace{.2cm}
\caption{Times (in seconds) for one inversion on the quenched $16^4$
 configuration at  $\beta = 6.0$, run on 16 processors of ALiCE.
\label{tab:times2.16}}
\vspace{.2cm}
\end{small}
\end{table}

\subsection{Chiral projection.}\label{chiralproj_sec}

Based on investigations of the Schwinger model,
the authors of \cite{BODE} suggested that it might be beneficial to
project the squared overlap operator to one chiral sector. We have
\begin{eqnarray*}
D_h^2 &=& D_+^2 + D_-^2;\nonumber\\
 D_{\pm}^2 &=& \frac{1}{2} D_h^2 \left(I \pm \gamma_5\right) \\
   &=& \frac{\rho^2+1}{2} \left(I \pm \gamma_5 \right) \pm  \frac{\rho}{2}\left(I \pm \gamma_5\right) \sign{Q}
        \left(I \pm \gamma_5\right).
\end{eqnarray*}

Because $[D_h^2,\gamma_5] = 0$, the eigenvalues\footnote{For a more
  detailed discussion of the eigenvalue spectrum of $D_u$ and $D_h$
  see ~\cite{vdEetal:03a}.}, $(\lambda^h_{j\pm})^2$ of $D_{\pm}^2$
are the same, except for the zero modes and their partners. If there
are no zero modes then
\[
\det D_u = \det D_+^2 = \det D_-^2,
\]
since the non-zero eigenvalues of $D_u$ are
\[
\lambda_{j\pm}^u = \frac{(\lambda^h_{j})^2}{2\rho} -\frac{1-\rho^2}{ 2\rho}
 \pm i\sqrt{(\lambda^h_{j})^2 -
 \left(\frac{(\lambda^h_j)^2}{2\rho} -\frac{1-\rho^2}{ 2\rho} \right)^2}.
\]
Zero modes can be treated exactly
at the end of the simulation by re-weighting the observables
according to $(2(1-\rho)/(1+\rho))^{|Q_f|N_f}$, where $Q_f$ is the fermionic
topological charge\footnote{An alternative is to introduce a
  second Metropolis step after a certain number of trajectories}. This means that
we can run $N_f=1$ simulations by projecting $\phi$ into one chiral sector
and running the HMC with $D_+^2$ rather than $D_h^2$. The calculation of
$D_{\pm}^2$  only requires one call to the sign function rather than two, so
in principle working in one chiral sector should run an $N_f=1$-simulation
in half the time it takes to run an $N_f = 2$-simulation without the
projection.

There are two advantages in using the chiral projection. Firstly, we can
work in the topological sector that contains no zero-modes, which means that
the inverse of the Dirac operator exists even at $\rho = 1$ ($\mu =
0$). The convergence of the inversion should be improved when the
fermion mass is of the same size as the smallest non-zero eigenvalues.
It is
however unlikely that on more realistic lattice sizes that we will be able to run at small enough masses to see
such an effect.

Secondly, and more importantly, it should
allow more frequent changes of the topological charge in the  chiral
sector opposite to the one we are working in, which means that better
samples in  configuration space and reduction of the
autocorrelation time can be achieved. This effect was seen in the Schwinger model
~\cite{BODE}, and our early results suggest that it is present
in four dimensions as well. In fact, HMC runs without chiral projection are very
resistant to changes in the topological charge.

However, there is also a disadvantage with this
method. At low fermion masses, the $Q_f=0$
configurations will dominate the statistical average. As a consequence
 $Q_f\neq 0$ configurations will turn out to be relatively
unimportant. Whether
these disadvantages outweigh the advantages is an open
question.
However, even if it is not advantageous to use chiral projection for
the up and down quark contributions to the determinant, it would
certainly be a useful tool if we wish to include a dynamical strange
quark in a $N_f=3$ simulation.
\begin{table}
\begin{small}
\begin{center}
\begin{tabular}{l l l l l l}
Method&Conf 1&Conf 2&Conf 3&Conf 4& Conf 5\\
\hline
CG            &631&564&623&635&641
\\
relCG         &399(1.58)&362(1.56)&399(1.56)&413(1.54)&398(1.61)
 \\
relGMRESR(CG) &190(3.32)&171(3.30)&180(3.46)&199(3.19)&181(3.54)
\\
\hline
\end{tabular}
\end{center}
\caption{Times (in seconds) for the inversion of chiral fermions on
  the $8^4,$ $\mu = 0.3$ ensemble, run on 16 processors of ALiCE
  \label{tab:times3}}
\end{small}
\end{table}

Table~\ref{tab:times3} summarises results for these chiral projection
computations with $D^2_+$ on the $8^4$ configurations with $\mu = 0.1$.  Since $D^2_+$ is hermitian positive definite, the unrelaxed and
relaxed computations were done with CG. The preconditioned variant took
relaxed GMRESR as the outer iteration with the CG iteration being the
preconditioner. There is, as expected, an
overall gain of a factor of approximately 2 for the chirally projected inversion, which
is due to the fact that we only need to call the sign function once for
each application of the squared overlap operator. This gain will not
appear in any HMC simulation, since we need to run two such
inversions to get a $N_F = 2$ ensemble.
However, single-flavour simulations come for free in this framework.
 Again, the gain from
relaxation and preconditioning is between 3 and 4, although the gain
from relaxation and preconditioning
is slightly smaller when chiral projection is used (compare Table
~\ref{tab:times3} and ~\ref{tab:times2}).

\subsection{The two pass SUMR inversion and propagator calculations.}

\begin{figure}[t]
\hfill \includegraphics[width =6.5cm,height =
  5cm]{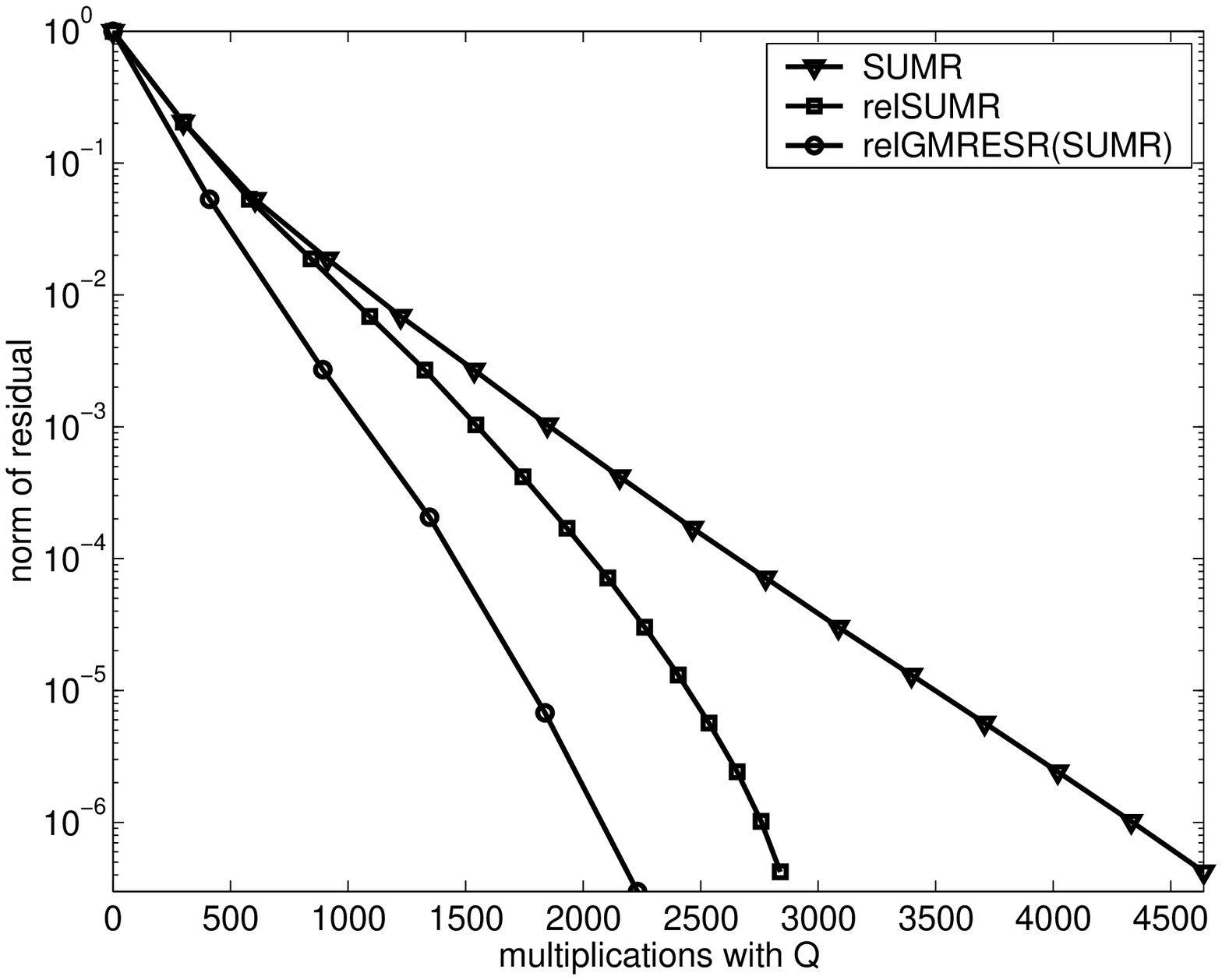}
 \mbox{} \hfill
\includegraphics[width =6.5cm,height =
  5cm]{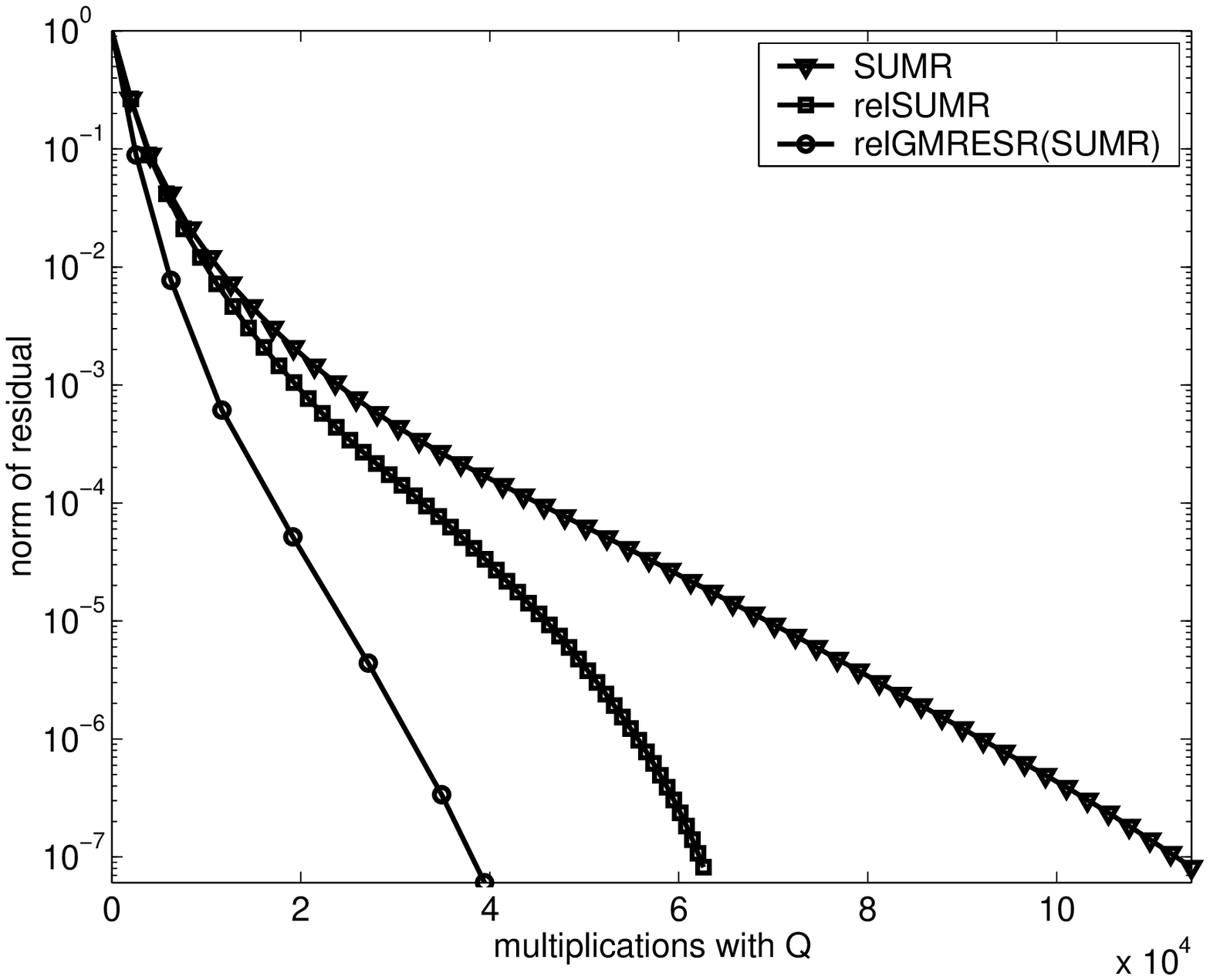} \hfill \mbox{} \vspace*{0.2cm}
  \\
  \mbox{}\hfill \begin{minipage}{4.5cm}first $4^4, \mu = 0.3$ configuration from Table~\ref{tab:times1b}
      \end{minipage}
  \hfill \begin{minipage}{4.5cm}
         third $8^4, \mu = 0.1$ configuration from Table~\ref{tab:times2}
		 \end{minipage}
  \hfill \mbox{}\\
\caption{Convergence history for unrelaxed SUMR, relaxed SUMR and  relaxed
  GMRESR(SUMR), two pass strategy for \eqnref{squared_eq}}
  \label{fig:sumr} \label{fig:sumr2}
\end{figure}

In a last series of investigations, we performed two pass computations
(see Section \ref{sec:KrylovMethods}) on
the $8^4$ and $4^4$ configurations. We now compare unrelaxed
SUMR, relaxed SUMR and relaxed GMRESR(SUMR), i.e.\ the
preconditioning is done with SUMR. As before, we required a fixed
accuracy $\xi_j = \xi= 0.1$ in the preconditioning iteration. As can be
seen in Figure \ref{fig:sumr} and Tables
\ref{tab:times1s} to \ref{tab:times2cs}, there is again an improvement of around
a factor of 3 to 4 when relaxation and preconditioning are used, and
again this factor increases as we decrease $\mu$, especially on the
larger lattices.  In
\cite{vdEetal:03a}, we predicted that two passes through SUMR should
theoretically take roughly the same time as one pass through CG, although in
numerical tests we found that the two pass method was slightly
slower. Comparing Tables  \ref{tab:times1s} to \ref{tab:times2cs} and
Tables \ref{tab:times1} to \ref{tab:times2b}, we can see that as we move
to larger lattices the unrelaxed two pass strategies do take
approximately the same time as the single CG
inversion. However, the gain in using the preconditioner is generally
larger for the SUMR inversion than for the CG inversion. If this trend
continues, then as we move to larger dynamical simulations, it might
be preferable to use the two pass strategy to calculate the HMC
fermionic force. Indeed, on our $16^4$ quenched configuration at the
lower masses, two SUMR inversions already improve upon one CG inversion when we use
the preconditioning technique.

To calculate the propagator we need to perform a single inversion of the
overlap operator using SUMR. The time needed for this calculation will
be half the time for the two pass inversions described in this section.

\begin{table}
\begin{small}
\begin{center}
\begin{tabular}{l l l l l l}
Method&Conf 1&Conf 2&Conf 3&Conf 4&Conf 5\\
\hline
SUMR           &68&59&59&66&63
\\
relSUMR        &48(1.42)&41(1.44)&41(1.44)&45(1.47)&44(1.43)
\\
relGMRESR(SUMR)&25(2.72)&24(2.46)&23(2.57)&32(2.06)&29(2.17)
\\
\hline
\end{tabular}
\\
$\mu = 0.3$
\bigskip

\begin{tabular}{l l l l l l}
Method&Conf 1&Conf 2&Conf 3&Conf 4&Conf 5\\
\hline
SUMR             &88&87&89&89&94
\\
relSUMR          &60(1.47)&61(1.43)&59(1.51)&64(1.39)&66(1.42)
\\
relGMRESR(SUMR)  &23(3.82)&21(4.14)&28(3.18)&30(2.97)&34(2.76)
\\
\hline
\end{tabular}
\\
$\mu = 0.1$
\end{center}
\vspace*{.2cm}
\caption{Times (in seconds)  for two SUMR inversions on the five $4^4$
 configurations at  $\beta = 5.4$, run on one processor of
  ALiCE.
\label{tab:times1bs}
\label{tab:times1s}
}
\vspace*{.2cm}
\end{small}
\end{table}

\begin{table}
\begin{small}
\begin{center}
\begin{tabular}{l l l l l l}
Method&Conf 1&Conf 2&Conf 3&Conf 4&Conf 5\\
\hline
SUMR           &1538&1181&1222&1288&1286
\\
relSUMR        &804(1.91)&748(1.58)&795(1.54)&818(1.57)&796(1.62)
\\
relGMRESR(SUMR)&383(4.02)&351(3.36)&383(3.19)&392(3.28)&382(3.37)
\\
\hline
\end{tabular}
\\
$\mu = 0.3$
\bigskip

\begin{tabular}{l l l l}
Method&Conf 1&Conf 2&Conf 3\\
\hline
SUMR             &2272&2685&2510
\\
relSUMR          &1695(1.34)&1661(1.62)&1500(1.67)
\\
relGMRESR(SUMR)  &674(3.38)&650(4.13)&576(4.36)
\\
\hline
\end{tabular}
\\
$\mu = 0.1$
\end{center}
\vspace*{.2cm}
\caption{Times (in seconds) for
two SUMR inversions on the $8^4$  configurations at $\beta = 5.6$, run on
16 processors of ALiCE.
\label{tab:times2s}
\label{tab:times2bs}
}
\vspace*{.2cm}
\end{small}
\end{table}

\begin{table}
\begin{small}
\begin{center}
\begin{tabular}{l l l l}
Method&$\mu = 0.03$&$\mu = 0.1$ & $\mu = 0.3$
\\
\hline
SUMR                &31550&8312&3200
\\
relSUMR                  &18840(1.87)&6038(1.38)&2656(1.20)
\\
relGMRESR(SUMR)  &5974(5.82)&2252(3.69)&1382(2.32)
\\
\hline
\end{tabular}
\end{center}
\vspace{.2cm}
\caption{Times (in seconds) for two SUMR inversions on the quenched $16^4$
 configuration, run on 16 processors of ALiCE.
\label{tab:times2cs}}
\vspace*{.2cm}
\end{small}
\end{table}

\section{Discussion}
Ginsparg-Wilson fermions, such as overlap fermions, offer an intriguing
possibility to overcome the bottleneck which affects
dynamical simulations with Wilson fermions at light quark masses. The lattice chiral symmetry
satisfied by the Ginsparg-Wilson fermions will also enable us to study aspects of
QCD such as chiral symmetry breaking and topology better than it is possible
with Wilson fermions.

The bottleneck of dynamical simulations with
Ginsparg-Wilson fermions is the computational time. In order to invert the
overlap operator in the course of the Monte-Carlo simulation, we need to run a
nested inversion, which means that overlap fermions are of the order $\mathcal{O}(100)$
times as expensive as Wilson fermions. The hope is that by improving
algorithmic techniques we can bring the
computational cost down to something manageable. In this paper, we have
studied two such techniques -- relaxation of the accuracy of the inner
inversion, and using a low accuracy approximation of the sign function
in a preconditioner. Considering the results on the $4^4$ lattices less typical, we conclude that these techniques lead to a factor of $3$ to $4$
improvement in the computational effort required to invert the
overlap operator. Improvements tend to be even better for more demanding computations, i.e. when $\mu$ becomes smaller. Our approach comes on top of the more classical eigenvector projection-techniques, to which it contributes its improvements in a multiplicative manner.

Using the techniques outlined in this paper (and anticipating further
improvements), we hope to be
able to run HMC algorithms using overlap fermions on moderate lattice sizes
on the next
generation of cluster computers. There are some subtleties when running
the HMC algorithm with overlap functions, the most notable being that the
derivative of the sign function generates a step function in the
fermionic force. We plan to discuss the Hybrid
Monte-Carlo algorithm in more detail and present some initial results
on small lattices in a subsequent paper.
\section{Acknowledgements}
N.C.
enjoys support from the EU Research and Training Network HPRN-CT-2000-00145
``Hadron Properties from Lattice QCD'', and EU Marie-Curie Grant
number MC-EIF-CT-2003-501467. A.F. and K.S. acknowledge support by DFG under grant Fr755/17-1. We thank Guido Arnold for his help in an early stage of this project.
 We are grateful to  Norbert Eicker and Boris Orth
for their help with the cluster computer ALiCE at Wuppertal University\@.

\begin{appendix}

\section{Definitions}\label{AppA}

The Wilson-Dirac matrix reads:
\begin{equation*}
M_{nm} = \delta_{nm} - \kappa D_W,
\label{eq:Diracmatrix}
\end{equation*}
where the hopping term is defined as
\begin{equation}
D_W = \sum_{\mu} (1-\gamma_{\mu})
U_{\mu}(n)\delta_{n,m-\mu} + (1+\gamma_{\mu})
U^{\dagger}_{\mu}(n-\mu)\delta_{n,m+\mu}.\label{HOPPING}
\end{equation}
$\kappa$ is the hopping parameter, which is defined as $\kappa =
1/(8-2m_0)$, where $m_0$ is the Wilson mass. The hermitian Euclidean $\gamma$ matrices satisfy
the anti-commutation relation $\left\{\gamma_i,\gamma_j\right\} =
2\delta_{ij}$ $i,j = 1,\ldots 4$. $\gamma_5$ is the product $\gamma_5 =
\gamma_1\gamma_2\gamma_3\gamma_4$, which means that
$\left\{\gamma_5,\gamma_{\mu}\right\} = 0$.
The hermitian form of the Wilson-Dirac matrix is given by multiplication of
$M$ with $\gamma_5$:
\begin{equation}
Q=\gamma_5\, M.
\label{HWD}
\end{equation}

\section{Massive Overlap Operator}

Following Neuberger ~\cite{hep-lat/9710089}, one can write the massive
overlap operator as
\begin{equation*}
  D_u(\mu) = c\left((1+\mu) + (1-\mu)\gamma_5\sign Q\right)\label{eq:overlapoperator}.
\end{equation*}
The normalisation $c$ can be absorbed into the fermion renormalisation, and
will not contribute to any physics. For convenience, we have set $c=1$.
Thus, the regularising parameter $\rho$ as defined in \eq{overlapmass:def} is
related to $\mu$ by
\begin{equation}
\rho= (1+\mu)/(1-\mu).
\label{REGULAR}
\end{equation}
The mass of the fermion is given by
\begin{equation*}
m_f = Z_m \frac{2\mu m_0}{(1-\mu)},
\end{equation*}
where $Z_m$ is a renormalisation factor.

Another form of the massive overlap operator, which sometimes
appears in the literature (e.g. in ~\cite{Chiu:2003iw}), is
\begin{equation*}
D_u = m + (m_0 - \frac{1}{2} m)(1 + \gamma_5\sign Q).
\end{equation*}
 This is equivalent to the
formula which we use, with $\mu = m/(2 m_0)$.

\section{Projection of low lying eigenvectors}\label{APPEIG}

It is advantageous to project out the low lying eigenvectors of the
Wilson operator $Q$ when calculating the sign function~\cite{xxxx} (see
Table \ref{tab:apptab1}).
Let the eigenvalues of $Q^2$ be contained in $[a,b]$. The smallest eigenvalues of $Q$ can be projected out
of the sign function and out of the multi-mass inversion used to
calculate the rational fraction~\cite{paper4}. We take the Zolotarev
approximation with respect to a domain
$[-\sqrt{b},-\sqrt{a'}]\cup[\sqrt{a'},\sqrt{b}]$ where $a' \geq
a$. Beforehand, we  compute a set $\Lambda$ of the $n_p$ smallest
eigenvalues of $Q$ and partition $\Lambda = \Lambda_1 \cup \Lambda_2$
where $\Lambda_2$ contains those eigenvalues which are larger in
modulus than $a'$. If $\psi^\lambda$ are the normalised eigenvectors of the eigenvalues $\lambda$
with respect to which we project, then

\begin{equation*}
\text{sign}(Q) x = \text{sign}(Q)\left(x-\sum_{\lambda \in \Lambda_1}
\psi^{\lambda}(\psi^{\lambda},x)\right) + \sum_{\lambda \in \Lambda_1}
\text{sign}(\lambda)\psi^{\lambda}(\psi^{\lambda},x)\label{eq:signproj}.
\end{equation*}
This shows that in order to compute $\sign{Q}x$ we need the Zolotarev approximation only on
the range $[-\sqrt{b},-\sqrt{a'}]\cup[\sqrt{a'},\sqrt{b}]$.

The projection approach in the subsequent multi-mass solver is to solve
\begin{eqnarray*}
\left( \frac{1}{{a'}}Q^2 + \tau_i \right)^{-1} y =& \left(\frac{1}{{a'}}Q^2 + \tau_i\right)^{-1}\left(y-\sum_{\lambda \in \Lambda_2}
\psi^{\lambda}(\psi^{\lambda},y)\right) +\nonumber\\& \sum_{\lambda \in \Lambda_2}
\frac{1}{\frac{1}{{a'}}\lambda^2 +
 \tau_i}\psi^{\lambda}(\psi^{\lambda},y)\label{eq:ratfracproj},
\end{eqnarray*}
where $y = x - \sum_{\lambda \in \Lambda_1} \psi^\lambda(\psi^\lambda,x)$.

This eigenvector projection improves the condition number of the
inversion, and therefore the CG method will converge
faster. Note that projecting all computed eigenvalues directly out of the sign function
would allow us to use a larger lower bound $a'$ for the Zolotarev
expansion
which will speed up the calculation further. However, this
comes with an additional cost when calculating the fermionic force,
and our preference is to only use this method for
exceptionally small eigenvalues.
Furthermore, in order to satisfy detailed balance, we need to use the
same Dirac operator throughout the calculation, i.e. we are forced to keep
$a'$ fixed.
\begin{table}
\begin{small}
\begin{tabular}{c c c c c}
$n_p$&Inversion&Calls to Wilson op.&Eigenvalue calculation&Total time\\
\hline
1&9144&1032172&0&9144
\\
10&1269&189514&111&1380
\\
20&796&112862&118&914
\\
30&568&78548&172&740
\\
40&459&63566&274&733
\\
50&387&52758&361&748
\\
60&340&45732&410&750
\end{tabular}
\caption{The times (in seconds) needed to calculate one relGMRESR(CG) inversion of
the overlap operator, and to calculate $n_p$ eigenvalues of the Wilson
operator for different values of $n_p$, on the $8^4$ configuration 1
with $\mu = 0.1$.
}\label{tab:apptab1}
\end{small}
\end{table}

\begin{table}
\begin{small}
\begin{tabular}{c c c c c}
$n_p$&Inversion&Calls to Wilson op.&Eigenvalue calculation&Total time\\
\hline
1&131&13112&18&149
\\
10&30&4860&14&44
\\
20&24&3532&22&46
\\
30&19&2874&31&50
\\
40&17&2474&60&77
\end{tabular}
\caption{The times (in seconds) needed to calculate one relGMRESR(CG) inversion of
the overlap operator, and to calculate $n_p$ eigenvalues of the Wilson
operator for different values of $n_p$, on the $4^4$ configuration
1, with $\mu = 0.1$.}\label{tab:apptab2}
\end{small}
\end{table}
The eigenvectors have to be calculated every time the gauge field is updated. In an HMC algorithm
this means that the time taken for each micro-canonical step is the
sum of the time taken for the calculation of the eigenvectors and the
time needed for the inversion of the overlap operator (the calculation of the remainder of the
fermionic force is negligible). Some fine tuning of $n_p$, the number
of eigenvectors projected out, is
therefore required. We used an Arnoldi algorithm with Chebyshev
improvement to calculate the lowest eigenvalues of the squared Wilson
operator~\cite{Neff}. From Tables \ref{tab:apptab1} and \ref{tab:apptab2},
we can see that there is a factor of 3 gain in using the eigenvalue
projection on the smaller lattices, and there is a large
factor of 12 on the
larger lattices.

\section{Hybrid Monte-Carlo with Overlap Fermions}\label{APPENDIXHMC}

\subsection{The Zolotarev approximation.}
 We
approximate the action of the matrix sign function $\sign{Q} = \gamma_5M (M^{\dagger}
M)^{-\frac{1}{2}}$ on a vector $y$
using the Zolotarev rational approximation. This ($l_\infty$) best approximation $r(\lambda)$
to $\sign{\lambda}$ on $[-\sqrt{b},-\sqrt{a}] \cup [\sqrt{a},\sqrt{b}]$ is given as\footnote{An alternative form of the Zolotarev
expansion can be found in ~\cite{LIUOVERLAP}.} \cite{PPo87}
\begin{equation}
\sign{\lambda} \sim r(\lambda) = \lambda \cdot D \cdot \frac{\prod_{j=1}^{N_Z-1}\left(\lambda^2 + c_{2j}\right)}{\prod_{j=1}^{N_Z}\left(\lambda^2 + c_{2j-1}\right)} , \label{eq:ratfrac}
\end{equation}
where  the coefficients are constructed using elliptic integrals as
\begin{eqnarray}
c_j =& \frac{\sn^2\left(j K/(2N_Z);\kappa\right)}{1-\sn^2\left(j
  K/(2N_Z);\kappa\right)}\nonumber\\
K = &\int_0^1\frac{dt}{\sqrt{(1-t^2)(1-\kappa^2 t^2)}}\nonumber\\
\kappa =&\sqrt{1-\frac{a}{b}}. \nonumber
\end{eqnarray}
$D$ is uniquely defined via the relation
\begin{equation}
\max_{\lambda \in [\sqrt{a},\sqrt{b}]}\left(1-\sqrt{\lambda}r(\lambda)\right) = -\min_{\lambda \in [\sqrt{a},\sqrt{b}]}\left(1-\sqrt{\lambda} r(\lambda)\right). \nonumber
\end{equation}
The rational function $r(\lambda)$ in \eqnref{eq:ratfrac} can equivalently be represented by
its  partial fraction expansion
\begin{eqnarray*}
  r(\lambda) =& \frac{1}{\sqrt{a}} \cdot \lambda \cdot \sum_{j =
    1}^{N_Z} \omega_{j} \left(\frac{1}{a}\lambda^2  + \tau_{j}\right)^{-1}
\end{eqnarray*}
where
\begin{eqnarray*}
 \omega_j &= &\frac{\prod_{k=1}^{n-1}(c_{2j-1} -
c_{2k})}{\prod_{k\neq j,k =
    1}^{k = n}(c_{2j-1} - c_{2k-1})} \\
\tau_j &=&c_{2j - 1}
\end{eqnarray*}
Using this representation, we approximate the action of the matrix sign function on a vector $y$ as
\begin{eqnarray*}
  \sign Q y \sim & \frac{1}{\sqrt{a}}Q \sum_{j =
    1}^{N_Z} \omega_{j} A_jy, \quad \mbox{ where $A_j =
    \left( \frac{1}{a}Q^2  + \tau_{j} I \right)^{-1}$}.
\end{eqnarray*}

Herein, the $ y^j = A_jy$ are calculated using the multi-shift CG method \cite{FM99,Je96}
as a multi-mass inverter for the systems $\left( \frac{1}{a}Q^2  + \tau_{j} I \right) y^j = y$.
For the outer iteration we set the
order of the Zolotarev expansion to be $N_Z=25$, which gave the sign
function accurate to machine precision when the multi-mass solver was
calculated to machine precision. When we needed less precision in the relaxed methods, we stopped
the multi-mass solver earlier, see \cite{EFL02}. When evaluating the sign function
in a preconditioner, where we only required an accuracy of $10^{-1}$, we could reduce the
order of the expansion to $N_Z = 5$.

\subsection{The fermionic force.}

The fermionic part of the Hybrid Monte-Carlo action is given by
\begin{eqnarray*}
S_{pf} =& \phi^{\dagger}X;\qquad &X = (H_N)^{-2}\phi.
\end{eqnarray*}
The fermionic force needed for the Hybrid Monte-Carlo algorithm at a
lattice site $x$ and direction $\mu$ is
\begin{eqnarray*}
F_{\mu}(x)
=&(1-\mu^2)\left(X^{\dagger}\gamma_5\right)_n\left(F^R_{\mu,nm}(b) +
F^P_{\mu,nm}(x) + F^S_{\mu,nm}(x)\right)X_m\nonumber\\
&(1-\mu^2)X^{\dagger}_n\left(F^R_{\mu,nm}(x) +
F^P_{\mu,nm}(x) +
F^S_{\mu,nm}(x)\right)\left(\gamma_5X\right)_m\nonumber\\
F^R_{\mu,nm}(x) =& \kappa \frac{1}{\sqrt{a}}
\omega_{\eta}A^{k}_{nb}\left[\frac{1}{a}Q_{bx}\gamma_5(1-\gamma_{\mu})\delta_{x+\mu,c}Q_{cd}
  - \right.\nonumber\\
&\left.\tau_{k}\gamma_5(1-\gamma_{\mu})\delta_{b,x}\delta_{x+\mu,d}\right]A^{k}_{de}\left(1-\ketbra\right)_{em}\nonumber\\
F^P_{\mu,nm}(x) =&-\kappa  \left(\frac{1}{\sqrt{a}}\frac{Q\omega_{k}}{\frac{1}{a}Q^2
  +\tau_{k}}\right)_{nb}P_{\lambda
  bx}\gamma_5(1-\gamma_{\mu})\delta_{x-e_{\mu},c}\left(\ketbra\right)_{cm}\nonumber\\
&-\kappa  \left(\frac{1}{\sqrt{a}}\frac{\frac{1}{a}Q\omega_{k}}{Q^2
  +\tau_{k}}\right)_{na}\left(\ketbra\right)_{bx}\gamma_5(1-\gamma_{\mu})\delta_{x,c+e_{\mu}}P_{\lambda
  cm}\nonumber\\
F^S_{\mu,nm}(x) =&\kappa P_{\lambda
  nx}\gamma_5\sign{\lambdabeta}(1-\gamma_{\mu})\delta_{x,c+e_{\mu}}\left(\ketbra\right)_{cm}+\nonumber\\
&\kappa
\left(\ketbra\right)_{nx}\gamma_5(1-\gamma_{\mu})\delta_{x,c+e_{\mu}}\epsilon(\lambdabeta)P_{\lambda
  cm}-\nonumber\\
&\left(\ketbra\right)_{nm}
\frac{d}{d\lambda}\sign{\lambdabeta}\bra{\psibeta}_x\gamma_5(1-{\gamma_{\mu}})\delta_{x,c+e_{\mu}}\ket{\psibeta}_c\nonumber\\
P_{\lambda} =& (1-\ketbra)(Q-\lambdabeta)^{-1}(1-\ketbra).
\end{eqnarray*}
We assume summations over all repeated indices, including 
sums over all the projected eigenvectors. Note that the
fermionic force contains a delta function in the smallest eigenvalue. This means
that if the smallest eigenvalue changes sign during the molecular
dynamics of the Hybrid Monte-Carlo, then some care needs to be taken
when calculating the fermionic force. We will discuss
this matter fully, and present our solution to the problem, in a
future publication~\cite{paper4} (see also~\cite{FODOR}).

In order to calculate the fermionic force, we need to perform two
multi-mass inversions of the Wilson operator and one inversion of the
squared overlap operator (e.g. by using relGMRESR(CG)). As discussed
during this paper, it is this
second step which is time-consuming.
We also need to calculate $S_{pf}$ during the
Monte-Carlo process, and for this we require just a single inversion of the
overlap operator
(e.g. by using relGMRESR(SUMR)).

\end{appendix}

\bibliographystyle{amsplain}
\bibliography{paper}

\end{document}